# Title Page

# Title: An early shutdown circuit for power reduction in high-precision dynamic comparators

Author one: Nima Shahpari

Department of Electrical Engineering, University of Isfahan, Isfahan, Iran.

Email: n.shahpari@eng.ui.ac.ir

Author two (Corresponding author): Mehdi Habibi

Department of Electrical Engineering, University of Isfahan, Isfahan, Iran.

Email: mhabibi@eng.ui.ac.ir

Author three: Piero Malcovati

Department of Electrical, Computer and Biomedical Engineering, University of Pavia, Pavia, Italy.

Email: piero.malcovati@unipv.it

# An early shutdown circuit for power reduction in high-precision dynamic comparators

N. Shahpari[1], M. Habibi[1*], P. Malcovati[2]

[1.] Dept. of Electrical Engineering, University of Isfahan, Isfahan, Iran.
[2.] Dept. of Electrical, Computer and Biomedical Engineering, University of Pavia, Pavia, Italy.

Abstract- Dynamic comparators are an essential part of low-power analog to digital converters (ADCs) and are referred to as one of the most important building blocks in mixed mode circuits. The power consumption and accuracy of dynamic comparators directly affects the overall power consumption and effective number of bits of the ADC. In this paper, an early shutdown approach is proposed to deactivate the first stage preamplifier at the suitable time. Furthermore, a time domain offset cancellation technique is incorporated to reduce offset effects. With the proposed method power consumption can be reduced in low power high precision dynamic comparators. The proposed method has been simulated in a standard 0.18µm CMOS technology and the results confirm its effectiveness. The proposed circuit has the ability of reducing the power consumption by 21.7% in the worst case, while having little effect on the speed and accuracy in comparison with the conventional methods. The proposed comparator consumes only 47µW while operating at 500MHz. Furthermore, Monte Carlo evaluations showed that the standard deviation of the residual input referred offset was 620µV.

## 1. Introduction

Wireless sensor networks are used in many applications. They consist of several sensor nodes which are spread geographically so that a physical characteristic of the environment can be collected. The sensor nodes are either powered up with internal batteries or use energy harvesting techniques [1] for power-up. In both cases, the most fundamental requirement in their design is low power consumption.

Analog to digital converters (ADC) have a crucial role in most digital sensor readout circuits. Low-power consumption and medium to high precision features have made SAR ADCs an appropriate choice for WSNs [2]. SAR ADCs consist of different building blocks such as DAC, successive approximation register, digital control logic and comparator. The power consumption and precision of each block determines the overall power consumption and ENOB of the ADC [3],[4].

Among the building blocks of an ADC, the comparator has a significant importance [5]. Dynamic comparators, in contrast to static comparators, are widely used because of their ability to eliminate the static power and hence, lowering the overall power consumption [6].

In addition to low power consumption, the comparator precision plays an important role in ADC design. The precision of the comparator, along with other sub blocks, determines the effective number of bits and output linearity in ADC design [7]. One of the most important parameters for comparator performance is the input-referred offset. The effect of offset is even more severe in dynamic comparators due to parasitic capacitances [8].

The most common method for overcoming the offset issue, is using a pre amplifying stage [9],[10]. By using the pre amplifying stage, the input referred offset is divided by the pre amplifier's gain, and hence, the input referred offset is reduced. However, due to the static power consumption of the pre-amplifier, this strategy is known as an energy hungry method.

To reduce the static power dissipation of the pre amplifier, double tail comparators can be used. This method adds a clock stage to the preamplifier. The clock signal controls the tail transistors of the preamplifier. With this approach the comparator power consumption is reduced by cutting off the path between power supply and ground in both the precharge and comparison cycles [11], [12].

In low-power designs, even the dynamic power consumption may be excessive. In these applications, other methods have to be considered to lower the dynamic power. In the paper by [13], the input devices are bulk driven so that lower supply voltages can be implemented. However, since the MOSFET devices are operated in the weak inversion and the transconductance is low in this region, the speed of the transistors is compromised [14], [15]. In these cases, the response time of the comparator is relatively low.

In the paper by Lu et. al. [16], a low-voltage, high-precision technique is proposed for dynamic pre amplification. It uses a novel offset cancelation method to minimize the input referred offset. This design uses the double tail preamplifier to reduce the first stage power consumption and applies an all dynamic feedback loop for cancelling the input referred offset.

Some modifications have been proposed to further reduce the overall power consumption of the comparator designed by Lu et. al. [17]. This technique uses an innovative structure to minimize the preamplifier's power consumption by turning it off before the outputs reach their saturation levels. This technique exploits the output of the latch comparator to determine the time when the final decision is ready. Subsequently a feedback loop turns the preamplifier off, since its output does not significantly affect the result generated on the latch. In another work, the preamplifier's reset voltage is adjusted at $V_{dd}/2$ to reduce the time of decision making in the latch block [18] and subsequently decrease power usage.

In this study, an early shutdown method is proposed which exploits the first stage comparator output results to turn off the input devices. [19] has used a similar approach for power reduction, however, the reported results for power consumption are far greater than conventional approaches due to the need for large MOSFET devices. In our work, the need for large devices and subsequently higher power usage has been solved by using an additional offset cancellation phase and power-delay optimization.

In the proposed method it is shown that decrease in the device dimensions reduces the power consumption. However, the use of smaller devices increases the input referred offset. Thus to cancel the input referred offset a time domain offset cancellation technique is also exploited.

The rest of the presented paper is organized as follows: In the second section a description of the presented circuit is given and the design considerations. In the third section, results of evaluating the proposed scheme are presented and finally the last section brings a conclusion to the paper.

## 2. Proposed Method

As illustrated in Figure 1, the proposed method exploits an early shutdown structure to reduce the power consumption of the preamplifier stage. This structure senses the preamplifier's output signal and uses it to control the tail current of the preamplifier stage. Using the output signals of the second stage latch (Vo+ and Vo-) can also be an option if designed properly as proposed in [17]. However, since the outputs of the first stage preamplifier (O+ and O-) arrive earlier, in this work they are used to shut down the preamplifier circuit at an optimal time and hence reduce the overall power consumption.

Two complementary dynamic voltage comparators are exploited for sensing the first stage output voltages $Out^-$ and $Out^+$. If the voltage of either $Out^-$ or $Out^+$crosses the reference voltage $V_{ref}$, the tail current source of the preamplifier has to turn off. The value of $V_{ref}$, is equal to the voltage at which the latch makes the final decision. When $Out^-$ or $Out^+$crosses $V_{ref}$, the output voltage of the first stage has insignificant effect on the latch result. Therefore, the voltage comparator's state can be changed by deactivating the tail current source. The correct adjustment of the deactivation point is important here since it should be chosen such that the preamplifier is turned off as early as possible while still allowing the latch stage to reach the decision making point. Failure to meet these requirements will result in higher power dissipation or longer delay time.

Some considerations were necessary for implementing the idea with low power consumption. In order to implement the voltage comparators which compares the $Out^-$ and $Out^+$ signals to $V_{ref}$, a pair of dynamic digital voltage buffers (DDVB) are suggested. Instead of comparing the $Out^-$ and $Out^+$ signals with $V_{ref}$, their delay is designed so that they turn on when the input signals reach $V_{ref}$. In order to implement each DDVB, two stages of dynamic inverter cells are used. Figure 1 shows the overall circuit designed in this study. The circuit behavior in the precharge and comparison cycles is discussed as follows.

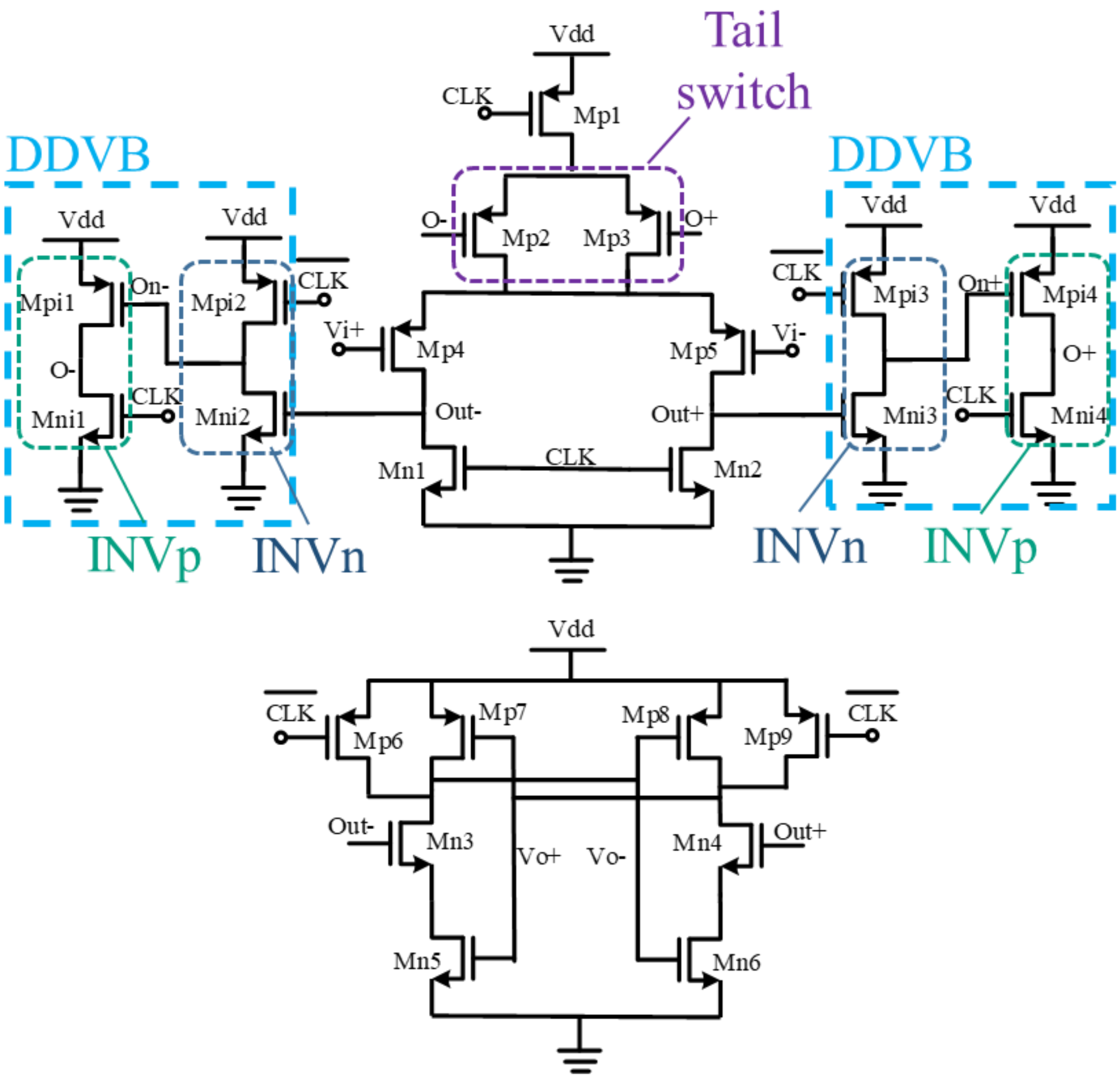

*Figure 1. Proposed comparator circuit*

a. Precharge cycle

In the precharge phase the positive feedback loop in the latch circuit is opened and all high impedance nodes are precharged to a specific voltage level. The precharge phase starts with the high state of clock signal (CLK=1). The high state of CLK, turns off the tail transistor of the preamplifier stage, Mp1, while Mn1 and Mn2 devices are turned on. Mn1 and Mn2 reset the output voltages of the first stage ($\mathrm{Out}^-$ and $\mathrm{Out}^+$) to ground. Since Mp1 is turned off, the static current from the source voltage to ground will be insignificant. The low voltages of $Out^-$ and $Out^+$, are fed into the second stage.

The high level of CLK will also precharge the DDVB inverters. This along with the low state of $Out^-$ and $Out^+$results in the reset of $O^-$ and $O^+$ nodes to ground. Therefore, Mni2, Mni3, Mpi1 and Mpi4 are turned off and the inverters do not consume any static power. The low level of $O^-$ and $O^+$ nodes, forces Mp2 and Mp3 in the triode region. This eliminates the effect of switches on the speed of the comparator since at the beginning of the comparing cycle the switches are fully on.

In the second stage, the low state of $\overline{CLK}$ signal will turn on Mp6 and Mp9 devices and the output voltages $Vo^-$ and $Vo^+$ will be set to $V_{dd}$. Furthermore, the low voltage level of $Out^-$ and $Out^+$ will place Mn3 and Mn4 devices in the off region and the path from the voltage supply to ground is cut.

b) Comparison cycle

The comparison cycle starts with the falling edge of the CLK signal. In this phase, transistors Mn1, Mn2, Mp6 and Mp9 are turned off and Mp1 is turned on. Subsequently the input devices start comparing the input voltages and the output voltages of the first stage $\mathrm{Out}^-$ and $\mathrm{Out}^+$ start to rise. At the beginning of the comparison phase, Mp2 and Mp3 are in the triode region, acting as an on switch. By proper design of the W/L of these transistors, their effect on the tail current and hence the speed of the preamplifier stage can be minimized.

It can be shown that in the comparison cycle, the current flowing through the input devices are dependent on the input voltages [20]. However, the maximum current is limited by the tail current. $Out^-$ and $Out^+$ voltages continue rising based on the fact that they are controlled by a constant current charging the load capacitors. When $Out^-$ and $Out^+$ reach the threshold voltages of Mn3 or Mn4 (because Mn5 and Mn6 are in deep triode region, their drain voltages can be neglected), the input devices of the latch stage will turn on.

The input devices of the latch block turn on in the saturation region due to the high drain source voltage and low drain current. When each of the input devices turns on, its current discharges the comparator output node $Vo^-$ or $Vo^+$. In turn, the decrease in $Vo^-$ and $Vo^+$, turns on Mp7 and (or) Mp8. The slope of $Vo^-$ and $Vo^+$depends on the difference between $Out^-$ and $Out^+$. In the subsequent “delay optimization” discussion, it is assumed that $Vi^+$ is larger than $Vi^-$. Therefore, $Out^-$ leads $Out^+$, and at the end of the decision making cycle, $Vo^+$ and $Vo^-$ will be high and low respectively.

*Power-Delay optimization*

The idea of the designed circuit is to cut off the drain current of the first input stage after decision making in the latch stage has ended so that, excessive current can be removed. The transient response of different nodes in the proposed design is shown in Figure 2.

As illustrated in Figure 2, the input devices are turned off when their output reaches $V_{ref}$. $V_{ref}$, as discussed before, is the minimum voltage needed for the latch circuit to complete the decision making process. If the first stage starts turning off before the decision making is completed, the speed of the latch stage will decrease and hence, the overall comparator delay will increase. Therefore, the time needed for the early shutdown structure to turn off the input stage ($t_{ESD}$ in Figure 2) should be approximately equal to the decision making time ($t_{DM}$ in Figure 2) in the latch block.

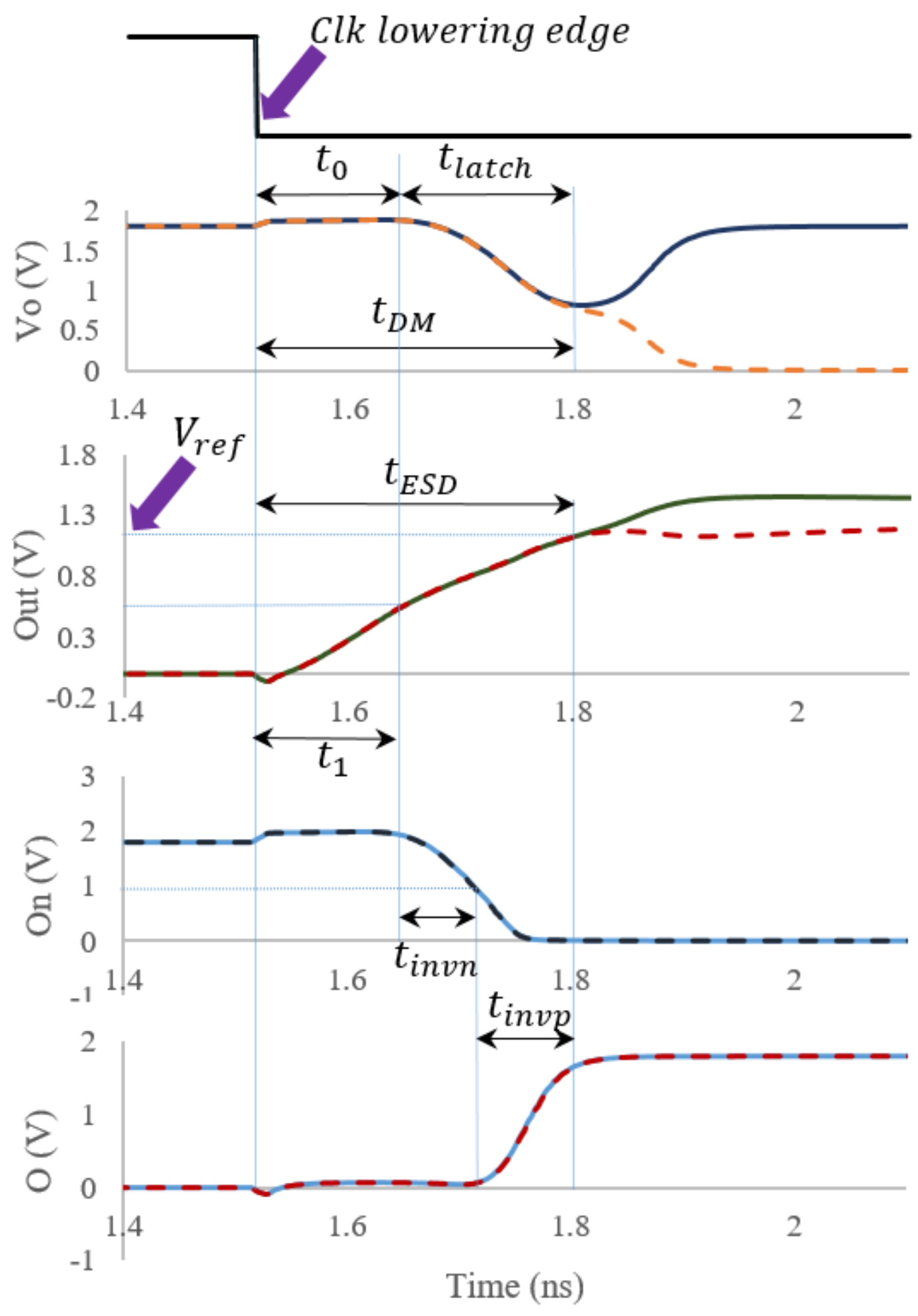

*Figure 2. Transient response of different nodes in the designed circuit.*

As illustrated in Figure 2, $t_{ESD} = t_1 + t_{invn} + t_{invp}$ where $t_1$, $t_{invn}$ and $t_{invp}$ are the time intervals for the first and second inverter to turn on, and the preamplifier stage to turn off, respectively. Assuming $Vi^+$ is larger than $Vi^-$, then $Out^-$ leads $Out^+$ and $Out^-$ first reaches the threshold voltage of Mni2. For $Out^+$ and $Out^-$ it can be written $Out^{+,-} = 1/(C_{n4,3} + C_{ni3,2}) \int I_{cp5,4} dt$, and based on the fact that $I_{cp5,4}$ does not change with time, $t_1$ can be extracted as follows:

$$t_1 = \frac{2Vthn(C_n+C_{ni})}{\beta_p(Vdd-V_i^{+,-}-Vthp)^2} \qquad \text{Eq. 1}$$

Due to symmetry it is assumed that, $C_{n3} = C_{n4} = C_n$, $C_{ni2} = C_{ni3} = C_{ni}$ and $\beta_{p4} = \beta_{p5} = \beta_p$. At $t_{inv1}$ the output of the first inverter starts falling from its initial $V_{dd}$ value. Since the inverter cells are dynamic with only one input device and charging one load capacitor ($C_{pi1} = C_{pi2} = C_{pi}$), $t_{invn}$ can be approximated as follows [21]:

$$t_{PHL} = t_{invn} = \frac{1.6C_{pi}}{\beta_n Vdd} \quad \text{Eq. 2}$$

where, $t_{PHL}$ is the high to low propagation delay of the $INV_n$. Similarly the delay time of $INV_p$ can be expressed as:

$$t_{PLH} = \frac{1.6C_{p3}}{\beta_p Vdd} \quad \text{Eq. 3}$$

where, $t_{PLH}$ is the low to high propagation delay of $INV_p$. Since $t_{invp}$ is defined as the time required for the tail switch to completely turn off, $t_{invp}$ can be estimated as $t_{invp} = \alpha \times t_{PLH}$ in which $\alpha$ is a multiplier greater than one (suggesting that it takes longer for the switch transistors to turn off than the output voltage to reach $V_{dd}/2$). According to Eq. 2 and Eq. 3, the delay in the inverter cells can be reduced by decreasing the load capacitance or increasing the W/L ratio of the active device of the inverter cell. The load capacitance of the first proposed dynamic inverter has been decreased by a factor of two by using only one input device dynamic inverter structure. From Eq. 1-3, the overall delay of the early shutdown structure $t_{ESD}$ can be estimated as:

$$t_{ESD} = \frac{2Vthn(C_n + C_{ni})}{\beta_p (Vdd - V_i^{+,-} - Vthp)^2} + \frac{1.6C_{pi}}{\beta_n Vdd} + \alpha \times \frac{1.6C_{p3}}{\beta_p Vdd}. \quad \text{Eq. 4}$$

As illustrated in Figure 2, the time interval for the latch stage to make the final decision can be written as $t_{DM} = t_0 + t_{latch}$, where, $t_0$ and $t_{latch}$ are the time needed for the first stage to turn Mn3 and Mn4 devices on and the time when the latch makes the final decision, respectively. $t_{DM}$ can also be regarded as the overall comparator delay. As shown in Figure 2, $t_0$ and $t_1$ are approximately the same, since both are the time intervals for the first stage preamplifier output to reach $V_{thn}$. Assuming that in the decision making phase the delay of the latch can be calculated in the same way as in an inverter cell (Due to the fact that only the NMOS devices contribute to the latch falling edge in the decision making time interval), $t_{latch}$ can be expressed as follows:

$$t_{latch} = \frac{1.6(C_{p8} + C_{n6})}{\beta_{n3} Vdd} \quad \text{Eq. 5}$$

Eq.8 is based on the fact that $V_o^-$ reaches the turning on point of the PMOS devices first (due to the assumption of $V_i^- < V_i^+$). Subsequently, the overall decision making time $t_{DM}$ is:

$$t_{DM} = \frac{2Vthn(C_n + C_{ni})}{\beta_p (Vdd - V_i^{+,-} - Vthp)^2} + \frac{1.6(C_{p8} + C_{n6})}{\beta_{n3} Vdd}. \quad \text{Eq. 6}$$

As mentioned earlier, $t_{DM}$ can approximately be considered as the overall comparator delay. Eq. 6 shows that implementing the proposed technique has only added the value of $C_{ni}$ in the first term of overall delay. $C_{ni}$ is the input capacitance of $INV_n$, which due to the implementation of one input transistor inverter, it is halved in comparison with a conventional CMOS inverter cells. In order to further minimize $C_{ni}$, Mni2 and Mni3 dimensions should also be set to minimum size.

As it is illustrated in Figure 2, the best time to cut off the preamplifier's tail current is when the decision making in the latch comparator is finalized. In fact, if the preamplifier turns off later than $t_{DM}$, more power will be consumed in the preamplifier stage. On the other hand, turning off the preamplifier sooner than $t_{DM}$ will increase the decision making time and hence the overall comparator delay will increase. Thus, considering the delay time and preamplifier power consumption, the optimum time for turning off the preamplifier is obtained when $t_{DM} \approx t_{ESD}$. Recalling that $t_0 \approx t_1$, from Eq. 4 and Eq. 6 we can write:

$$\frac{C_{pi}}{\beta_{ni}} + \alpha \times \frac{C_{p3}}{\beta_{pi}} \approx \frac{(C_{p8}+C_{n6})}{\beta_{n3}} \qquad \text{Eq. 7}$$

It was mentioned earlier that, because of comparator delay considerations, Mni2 and Mni3 should be minimum size and therefore, $\beta_{ni}$ will be minimized. Furthermore, the right-hand side in Eq. 7 has to be small enough because, based on Eq. 6, it forms an important part of the comparator delay $t_{DM}$. Therefore, Mn3 and Mn4 are designed with rather larger dimensions. Simplifying Eq. 7 with the device lengths set to minimum feature size and assuming similar NMOS and PMOS oxide capacitances, Eq. 8 is derived as follows:

$$\frac{W_{pi}}{\mu_n W_{ni}} + \alpha \times \frac{W_{p3}}{\mu_p W_{pi}} \approx \frac{(W_{p8}+W_{n6})}{\mu_n W_{n3}} \qquad \text{Eq. 8}$$

It was discussed that $W_{ni}$ should be chosen minimum to decrease the effect of early shutdown structure on the overall delay in the previous section. The width of other devices can be written as multiples of $W_{ni}$. Assuming $W_{n6} = W_{p8} = 2W_{n3}$, and $\mu_n \approx 2\mu_p$, Eq. 8 turns into:

$$\frac{x}{2} + \alpha \times \frac{y}{x} \approx 2 \quad : x \geq 1, y \geq 1 \qquad \text{Eq. 9}$$

where x and y are the $W_{pi}/W_{ni}$ and $W_{p3}/W_{ni}$ respectively and since $W_{ni}$ is chosen to be minimum size, x and y can be greater or equal to 1 (suggesting that $W_{pi}$ and $W_{p3}$ cannot be chosen below the minimum size). The closest solution for Eq. 9 can be obtained when $x = 1$ and y=1 ($W_{pi} = W_{ni}$).

Decreasing y, decreases the switch size which affects the maximum tail current of the preamplifier stage and reduces the speed of the comparator. Therefore, the parallel tail switch of Figure 1 is beneficial in this regard. This structure allows decreasing the dimension of the tail switches by half while the overall tail current is intact. Here, the load capacitors of $INV_p$ ($C_{p2}$ and $C_{p3}$) remain low while the tail current of the preamplifier is preserved and the delay of the comparator is not compromised. Furthermore, the parallel structure, in comparison with the series structure in the paper by khorami et. al. [17], has the benefit of imposing a symmetric load capacitor on the preceding block. The symmetric load capacitor decreases the propagation delay mismatch between the two output ports of the comparator. Based on these considerations, y was chosen as 1.5 so that, $W_{p2,3} = 1.5W_{ni}$.

In the proposed design the power consumptions of four inverter cells are added to the overall power and thus contribute to an overhead. The dynamic structure is proposed for the inverter cells to reduce the overhead power consumption. Since the transition occurs when the path from $V_{dd}$ to ground is opened, for a dynamic cell the short circuit power consumption is negligible. Besides, due to symmetry the power consumption of similar inverter cells is the same. As a result, the power consumption of the inverter cells can be written as $P_{inv} = 2 * (P_{invn} + P_{invp})$. For the power consumption of each inverter in the proposed dynamic inverter cells it can be shown that for a given $V_{dd}$ and operating frequency, the minimum power consumption can be achieved when $C_{pi}$ and $C_{p3,2}$ are minimized [21].

*Proposed offset cancellation scheme*

As shown in the previous section, to optimize the power-delay-product, the device dimensions have to be set to minimum. However, the minimum size input devices increases the input referred offset [18]. In order to reduce the input referred offset, an offset cancellation technique is proposed here.

The proposed time domain offset cancellation block diagram is illustrated in Figure 3. This technique exploits the time domain idea proposed by Judy et. al. [22]. However, in the proposed method the input referred offset is reduced by adjusting the body voltage of the input pairs instead of changing the output current of the preamplifier stage so that it does not add any additional loading to the preamplifier stage.

Similar to [16], the power consumption overhead of the proposed offset cancelation can be neglected since the cycle needs to be repeated at a rate much lower than the comparison clock frequency.

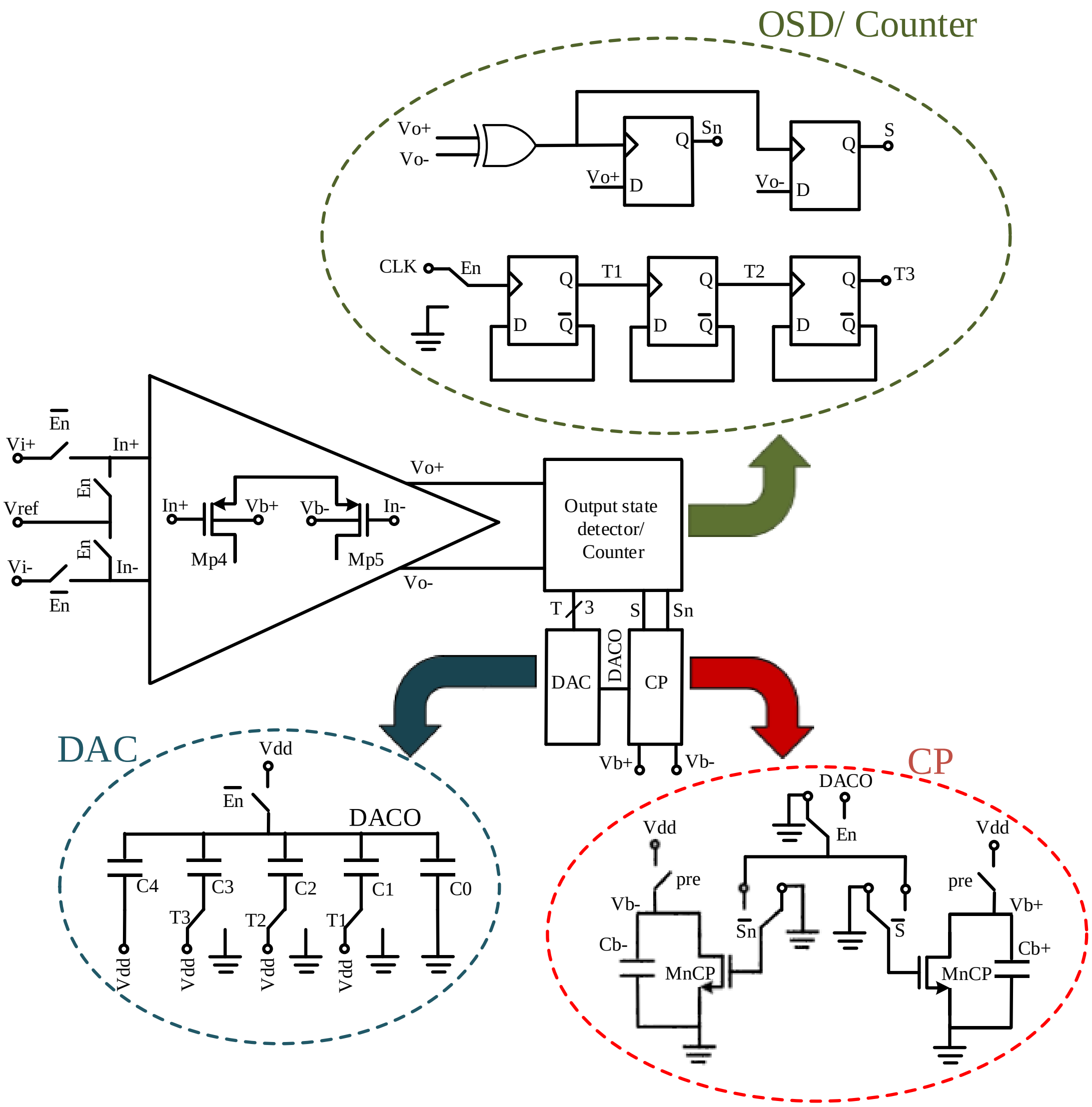

*Figure 3. The block diagram of the proposed time domain offset cancelation technique.*

As Shown in Figure 3, the proposed scheme uses the state of the output voltage, to tune the body voltage of the input devices. The change in the body voltages can alter the threshold of the input devices and hence, change their speed which will lead to input referred offset reduction [16],[20]. Figure 4 shows the transient behavior of different node voltages of the proposed scheme in the offset cancellation phase.

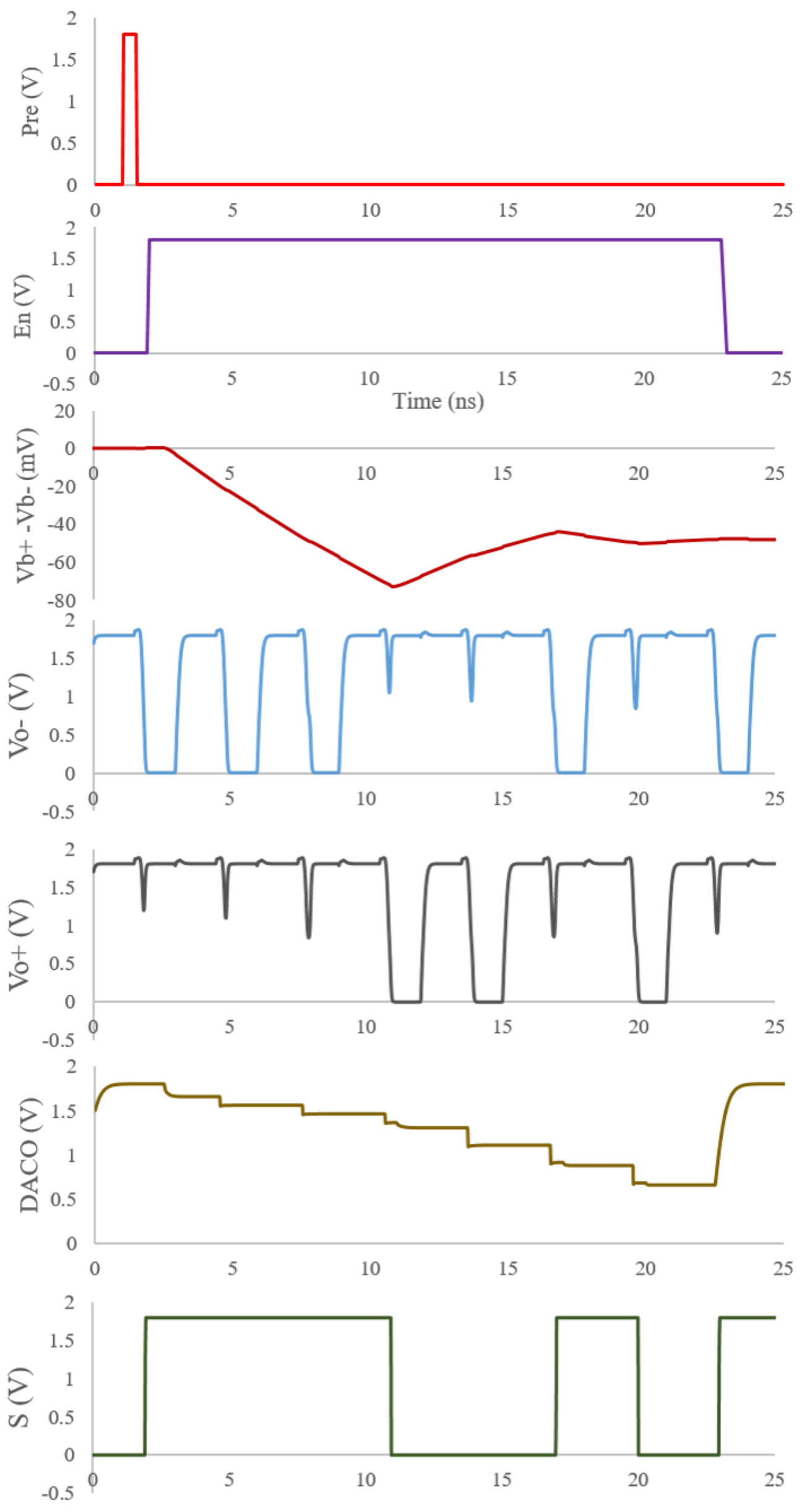

*Figure 4. Transient response of offset cancellation phase.*

Before the offset cancellation starts, the output voltages of $C_{b+}$ and $C_{b-}$ ($V_{b+}$ and $V_{b-}$) are primarily charged to $V_{dd}$ by the precharge signal (the first waveform of Figure 4), resulting in $V_{b+} - V_{b-} = 0$.

The offset cancellation phase starts with the high state of En signal (second waveform of Figure 4). Subsequently the offset cancellation sweep is commenced as shown in the third waveform of Figure 4.

Depending on the comparator output (the fourth and the fifth waveform of Figure 4), the charge pump block (CP) discharges one of the $C_b$ capacitors and hence, a difference between $Vb^+$ and $Vb^-$ ($\Delta V_b \neq 0$) is generated. With the high state of En signal the input ports will be connected to the same input reference voltage, $V_{ref}$. Here, $Vb^+$ and $Vb^-$ and DACO signals are disconnected from $V_{dd}$. The Counter block starts to count the clock signal cycles and will generate three bits of Tn for the digital to analog converter block (DAC).

The change in Tn binary codes will connect the lower terminal of the capacitors from $V_{dd}$ to ground. This structure makes the charge previously stored on C0 to be distributed with the selected Cn. Feeding the binary codes to DAC block, changes DACO from $V_{dd}$ in relevance to the clock cycle number as shown in the sixth waveform of Figure 4.

In the seventh waveform of Figure 4, the digital output state detector block (OSD) generates another signal (S) which corresponds to positive or negative input referred offset. The proper sign of $\Delta V_b$ is determined by the state of S. Assuming that the comparator has positive offset, $Vo^+$ is high and $Vo^-$ is low. Therefore, S signal changes to 1 indicating positive offset. Subsequently, CP block turns on and the gate terminal of MnCP is connected to DACO and this device turns ON. The activated MnCP starts to discharge $C_b$+ and $Vb^+$ decreases. The reduction in $Vb^+$ increases the speed of Mp4 by $I/C_bT$, where I is the current passing through MnCP, $C_b$ is the output capacitance of CP and T is the period of the operating cycle. The gate terminal of MnCN stays connected to the ground at this stage.

As mentioned, the reduction in $Vb^+$ increases the speed of Mp4 and at the next clock cycle Mp4 will be relatively faster in comparison with Mp5 transistor. The output state of the comparator indicates whether the sign of $\Delta V_b$ is set accordingly. If at the next cycle $Vo^+$ is still high and $Vo^-$ low, signal S will remain high and the previous cycle will be repeated for $Vb^+$. However, if $Vo^+$ becomes low, signal S state will change to low and the previous cycle will be commenced in the opposite direction i.e. on the body voltage of Mp5.

Using various gains for the offset cancellation phase decreases the offset cancellation settling time [22]. In order to change the loop gain of the offset cancellation phase, the gain of CP block in iterative cycles is changed. The counter and DAC blocks convert the offset cancellation cycle number to an analog voltage. The analog voltage is then provided as the input to the CP circuit. Eventually the CP gain is controlled by the number of offset cancellation cycles.

The proposed variable gain scheme is based on the assumption that the accumulation of $\Delta V_b$ after six clock cycles is large enough to compensate the highest mismatches. This means, in each offset cancellation phase (before six clock cycles end) $\Delta V_b$ will cross its final value. Therefore, in each offset cancellation cycle the gain has to be lowered to ensure $\Delta V_b$ approaches its final value.

## 3. Circuit evaluation

The proposed circuit was simulated in a standard 0.18µm CMOS technology. In order to make fair comparison, the paper by Lu et. al. [16] was simulated in the same technology and is referred to as the conventional method in the results. In all simulations, the following conditions are applied and referred to as typical conditions unless otherwise is mentioned:

- Supply voltage =1.8V
- Common mode voltage = $V_{dd}/2$
- Differential input voltage = 50mV
- Frequency=333MHz
- Temperature = 27°C
- Fabrication process was set to TT

In order to show the effect of device dimensions on the comparator performance, overall power consumption and delay vs. dimensions are presented in Figure 5. The parameters of the designed and conventional comparators are shown when the dimension of preamplifiers varies. Here, the width of the input devices Mp4 and Mp5 is W while the width of the tail device Mp1 is chosen as 2×W.

The change of comparator performance with the dimension change in $INV_n$ is shown in Figure 6. In order to keep the delay between the $INV_n$ and $INV_p$ constant, the same scaling was used for the two blocks. The effect of changing the width of Mpi1, Mni2, Mni3 and Mpi4 on the comparator parameters is also shown in Figure 7.

Based on the obtained results, the final optimized device dimensions of the proposed comparator, are shown in Table 1.

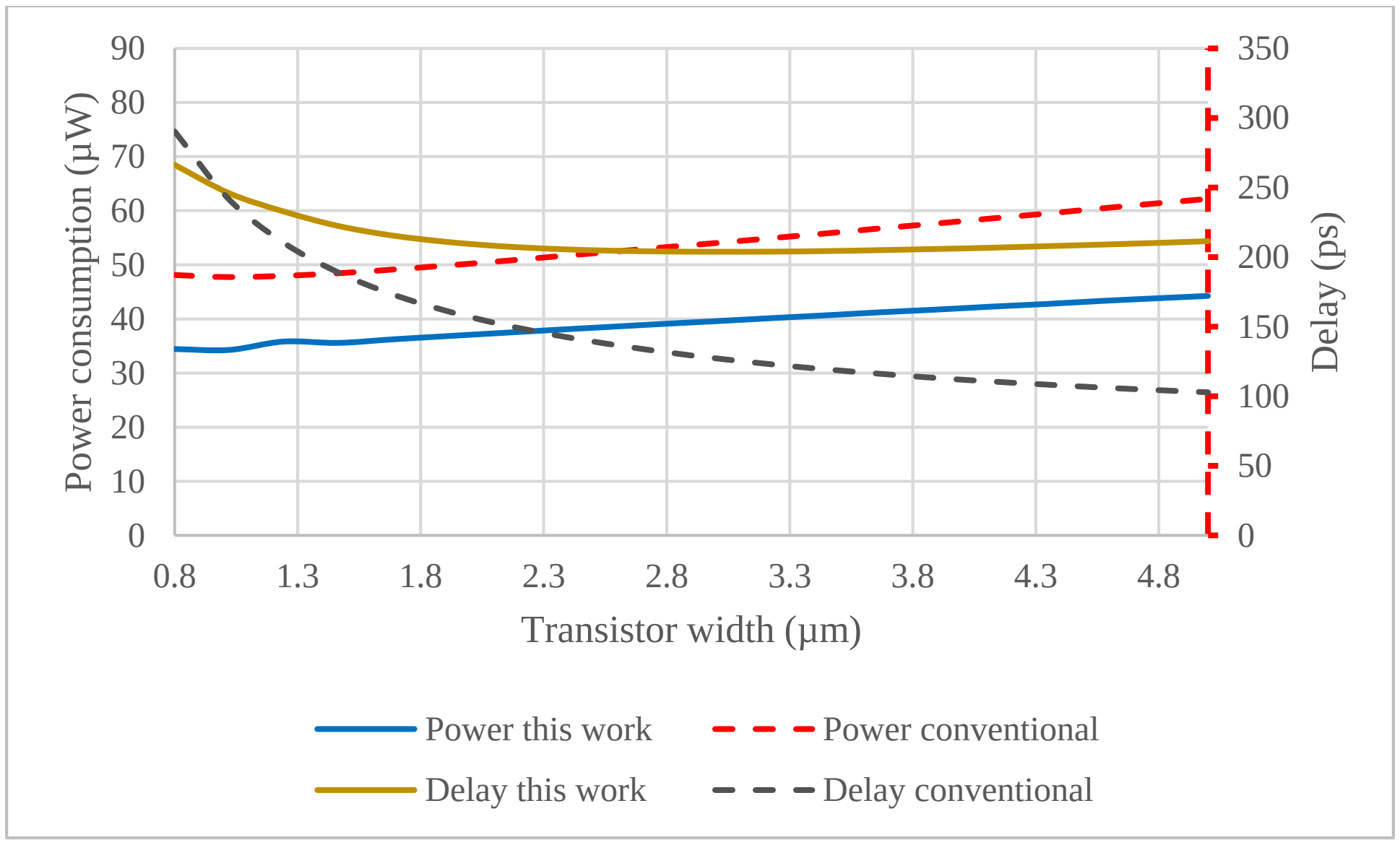

*Figure 5. Power consumption and delay of the overall comparator while the transistor width of the preamplifier changes for the proposed and conventional comparator.*

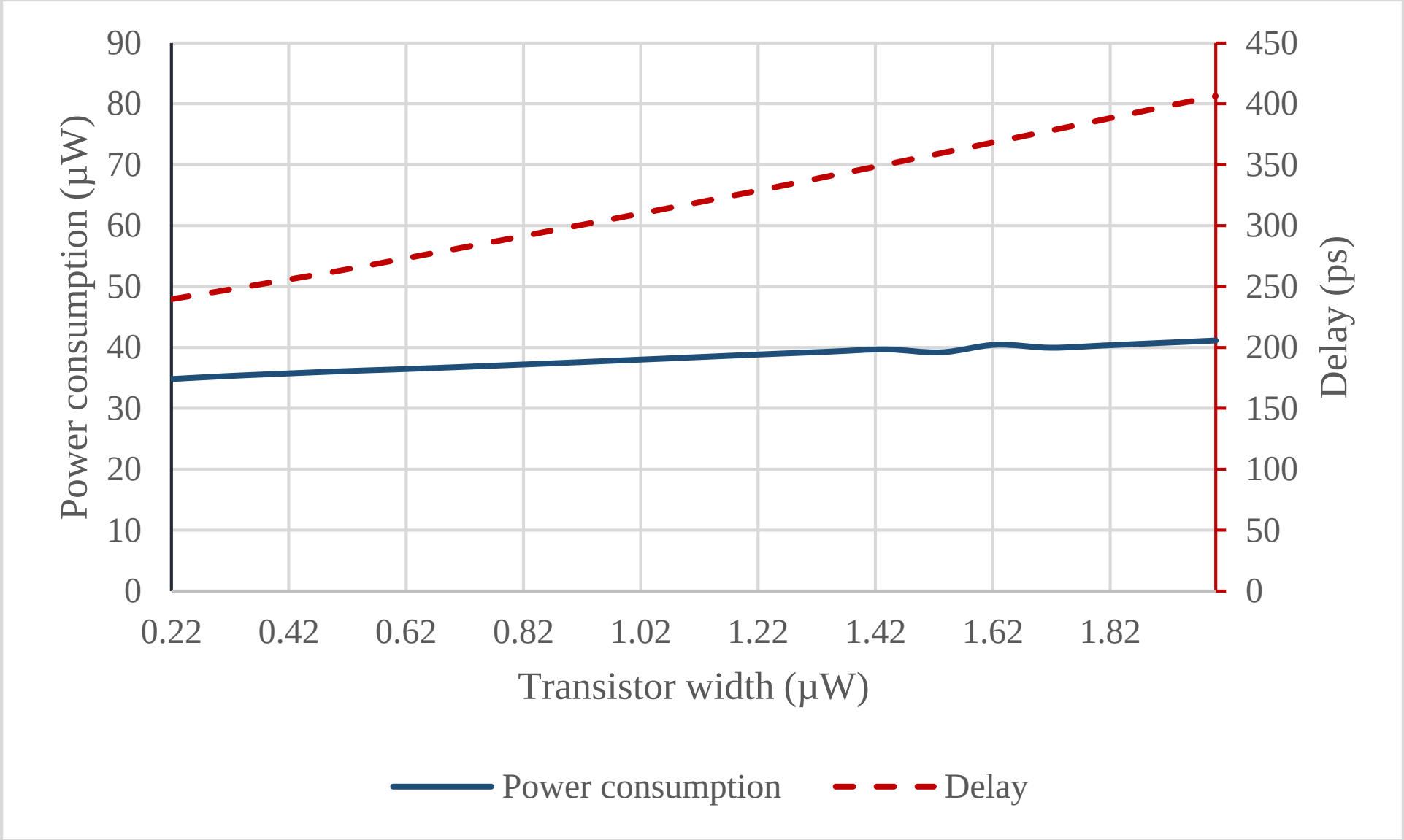

*Figure 6. Power consumption and delay of the overall comparator while the transistor width of the $INV_n$ change for the proposed.*

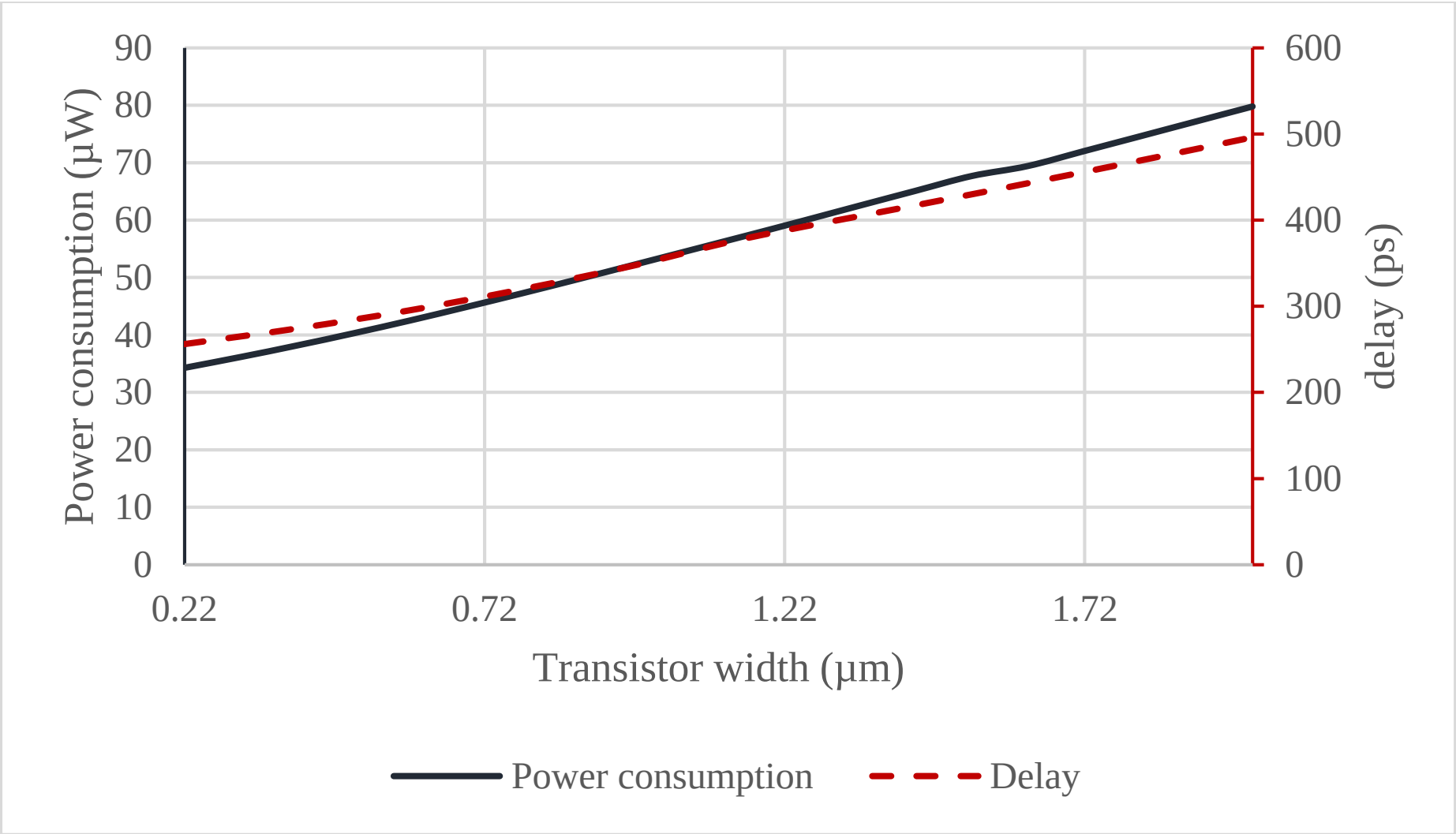

*Figure 7. Power consumption and delay of the overall comparator while the transistor width of the $INV_n$ and $INV_p$ change simultaneously, for the proposed.*

*Table 1. Dimensions of the MOSFETs of the proposed circuit*

| Transistor | W/L (μm) | Transistor | W/L (μm) |
|---|---|---|---|
| **Mp1** | 2/0.18 | **Mn1, Mn2** | 0.5/0.18 |
| **Mp2, Mp3** | 0.35/0.18 | **Mn3, Mn4** | 1/0.18 |
| **Mp4, Mp5** | 1.2/0.18 | **Mn5, Mn6** | 2/0.18 |
| **Mp7, Mp8** | 2/0.18 | **Mni1-Mni4** | 0.22/0.18 |
| **Mp6, Mp9** | 0.5/0.18 | | |
| **Mpi1-Mpi4** | 0.22/0.18 | | |

Because the aim of this study is to minimize the power consumption while preserving other specifications, to make fair comparison, the comparator delay is approximately set to the delay of the conventional design.

The delay of the comparator output versus the differential input voltage is shown in Figure 8 when the input differential mode voltage is varied from 1mV to 50mV. The delay is defined as the difference between the time at which the clock signal and the output signal reach 50% of their final values. Figure 9 shows the variation of the output delay as the input common mode voltage varies from rail to rail. In this figure, the vertical axis is logarithmically plotted so that the drastic increase in the delay of the comparator can be emphasized. To obtain delays near what is reported in previous architectures, the common mode input voltage of the presented method is considered below 1.1V.

The power consumption of the proposed and conventional comparator is compared in Figure 10. In this figure, the overall power consumption is shown as the input differential voltage varies from 1mV to 50mV. As expected, the overall power consumption decreases as the input differential voltage increases. This is due to the fact that for low input differential voltages, the comparator requires a longer time to decide the final state of its output.

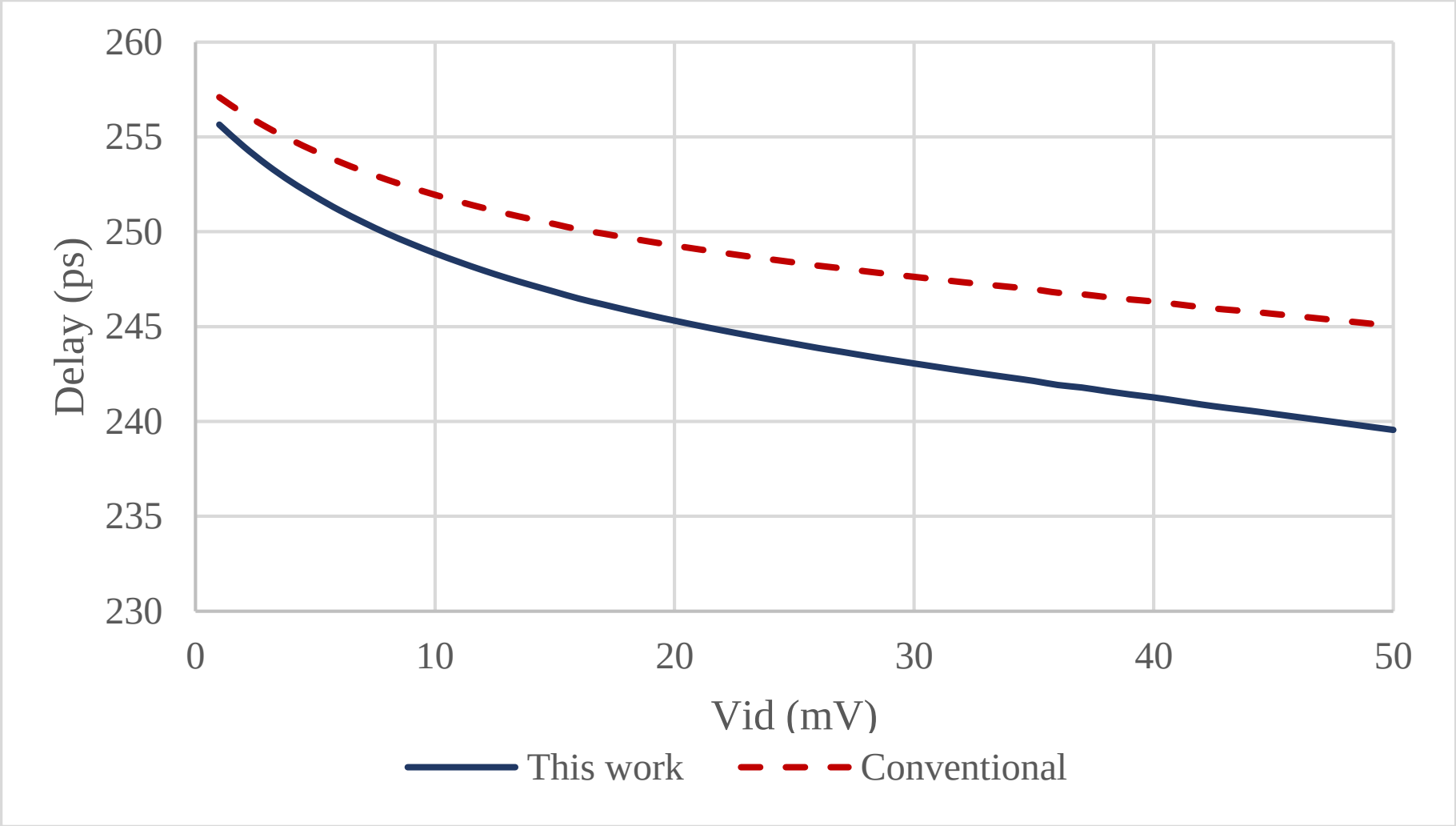

Figure 8. *Delay vs. input differential voltage of the proposed and conventional comparator.*

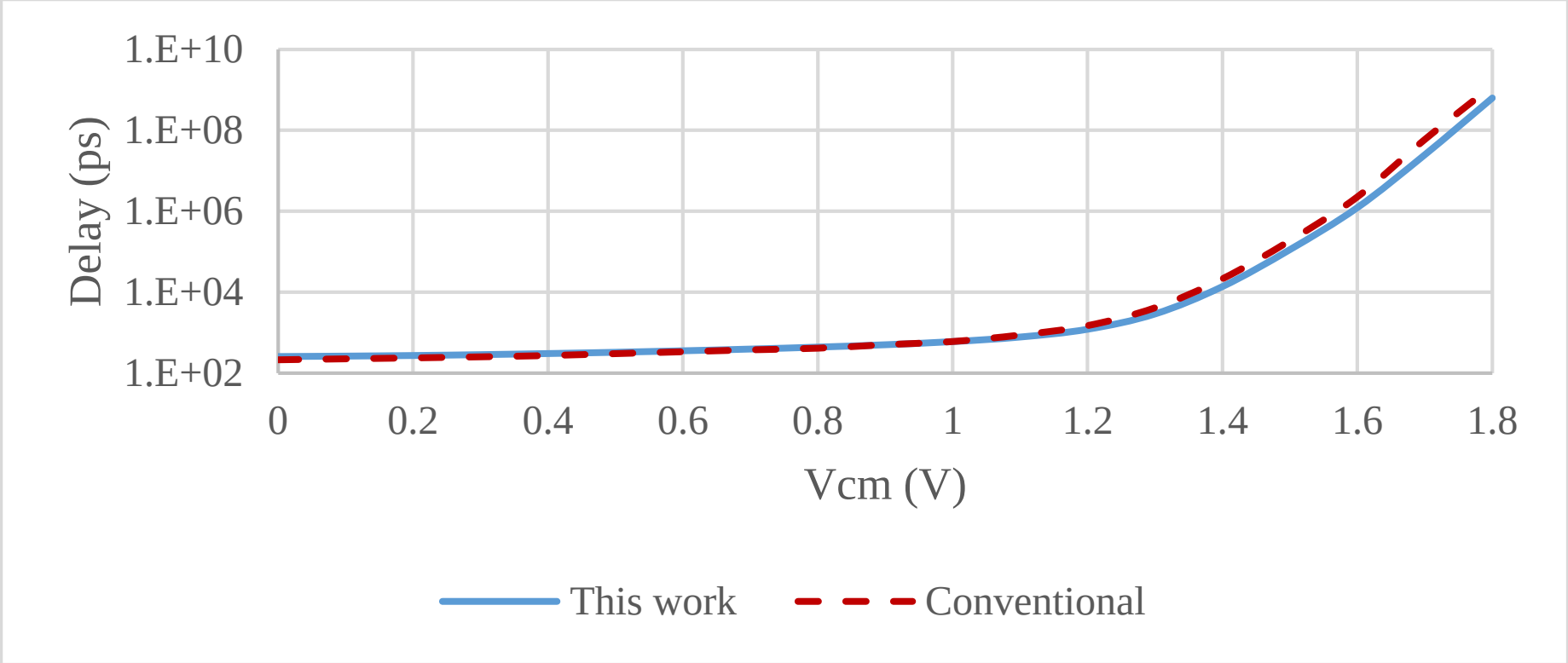

*Figure 9. Delay vs. input common mode voltage of the proposed and conventional comparator.*

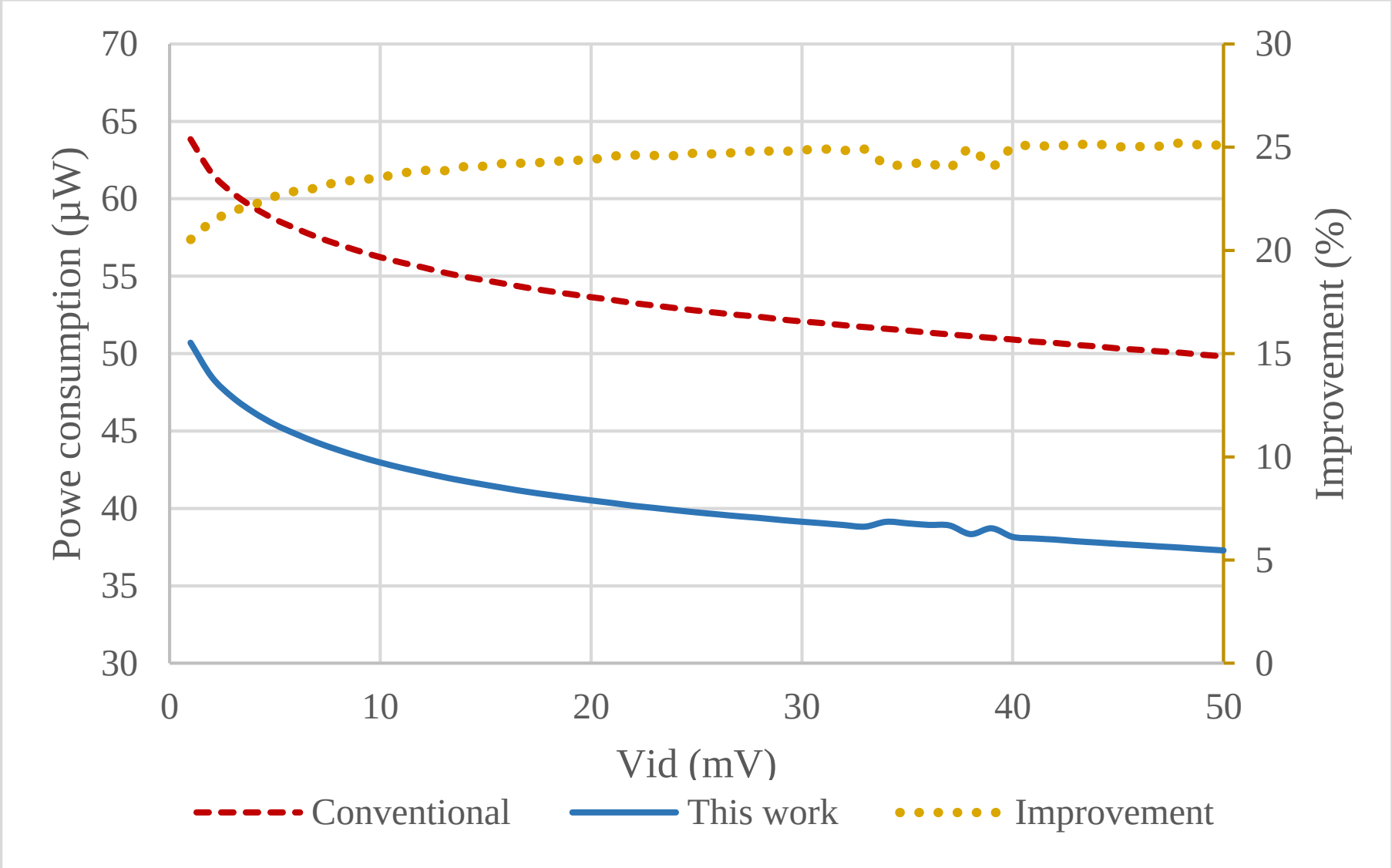

Figure 10 *Power consumption* vs. input differential voltage *of the proposed and conventional comparator.*

In order to evaluate the performance of the proposed comparator while the process, supply voltage and temperature (PVT) vary, different simulations were executed. The effect of PVT changes were studied on the comparator delay and power consumption.

The output delay of both the proposed structure and the conventional structure is shown in Figure 11 when the input differential voltage varies from 1mV to 50mV for four different process corners while other simulation parameters are set to typical. The results, as expected, show that for the FF and SS process corners, both comparators show lowest and highest delays respectively. Because of using PMOS input devices in the preamplifier stage, the FS corner shows rather slow behavior (even slower than the TT process shown in Figure 8).

The results of output delay are shown in Figure 12 when the input common mode voltage varies from 0.1V to 1.1V for four different process corners. The delay of the comparator rises with higher slope with the common mode voltage, when the process features slow PMOS devices. This is due to the fact that slower input devices for preamplifier stage tend to pass lower currents when the input voltages rises and therefore lower input common mode range is available in these process corners.

The effect of different process corners on the power vs. input common mode voltage is studied in Figure 13. This figure shows the effectiveness of the proposed structure in all the process corners. As expected, the faster processes have relatively higher power consumptions.

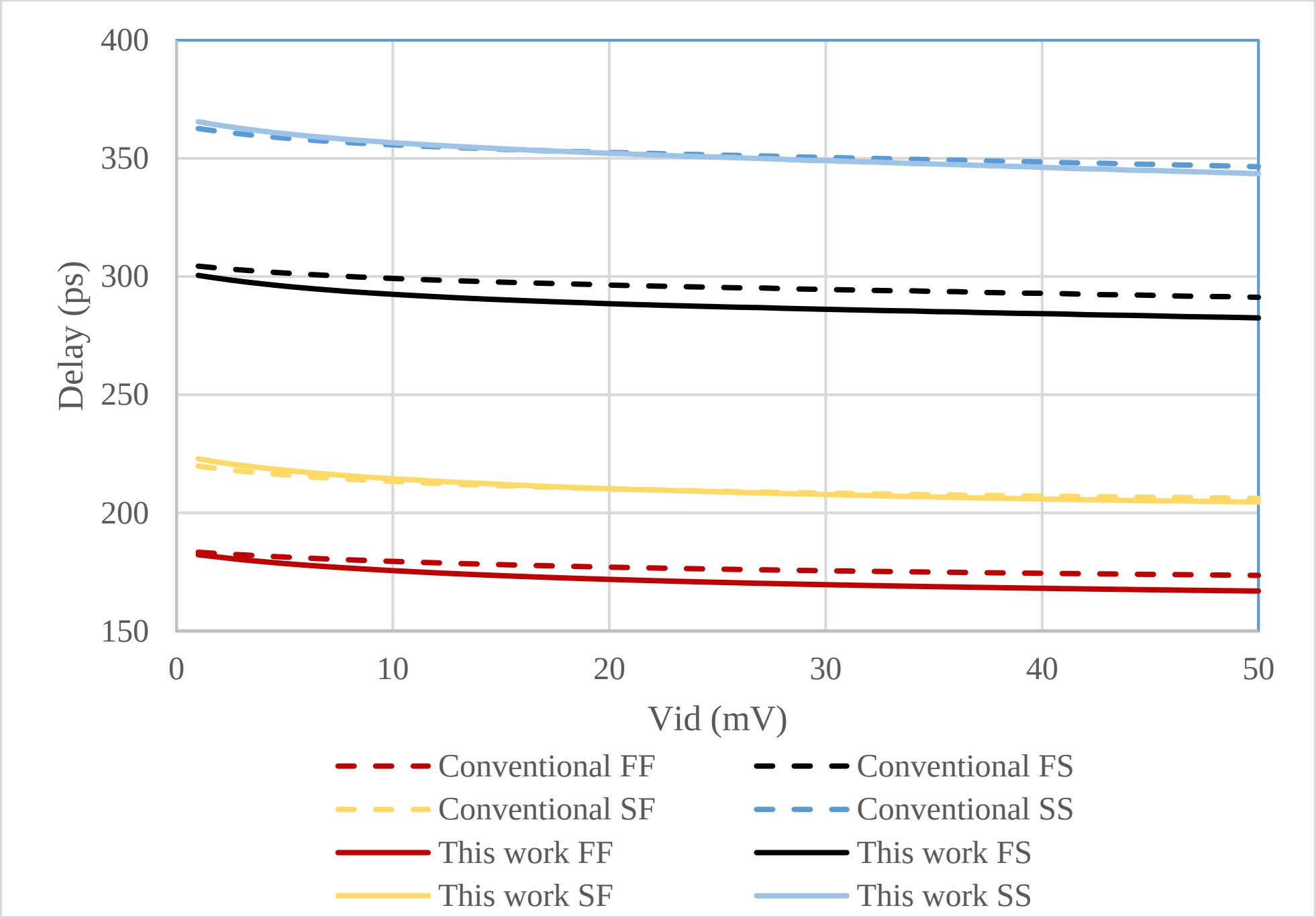

*Figure 11. Delay in different process corners vs. input differential voltage of the proposed and conventional comparator.*

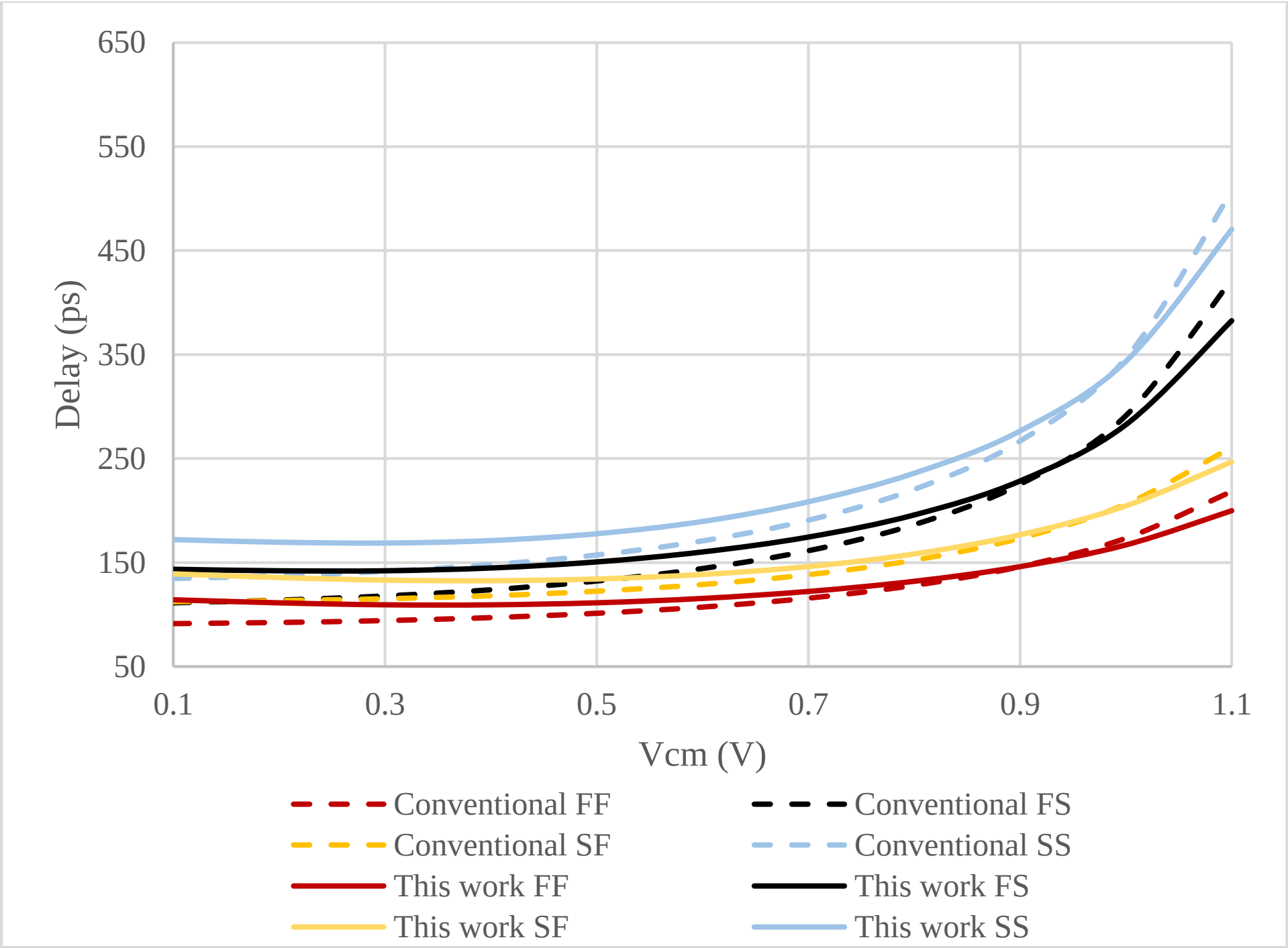

*Figure 12. Delay in different process corners vs input common mode voltage of the proposed and conventional comparator.*

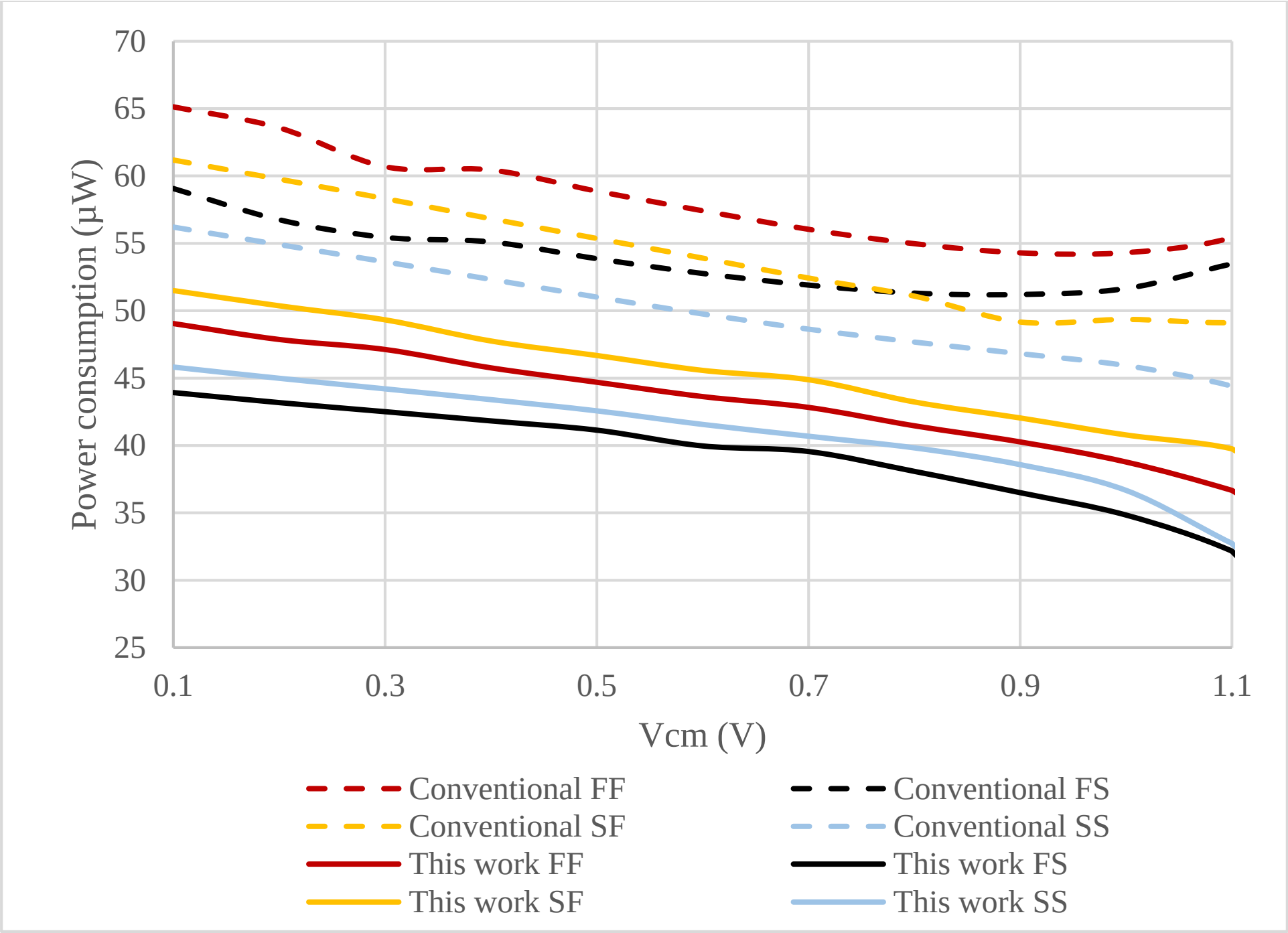

*Figure 13. Power consumption in different process corners vs input common mode voltage of the proposed and conventional comparator.*

Figure 14 and Figure 15 show the effect of power supply voltage on delay and power of the proposed and conventional comparator respectively. The percentage of power consumption improvement is also reported in Figure 15. The diagrams show that for both structures, higher voltage supply results in faster and more power hungry comparators. In the proposed structure, with higher supply voltages lower improvement in power consumption is obtained with respect to typical conditions.

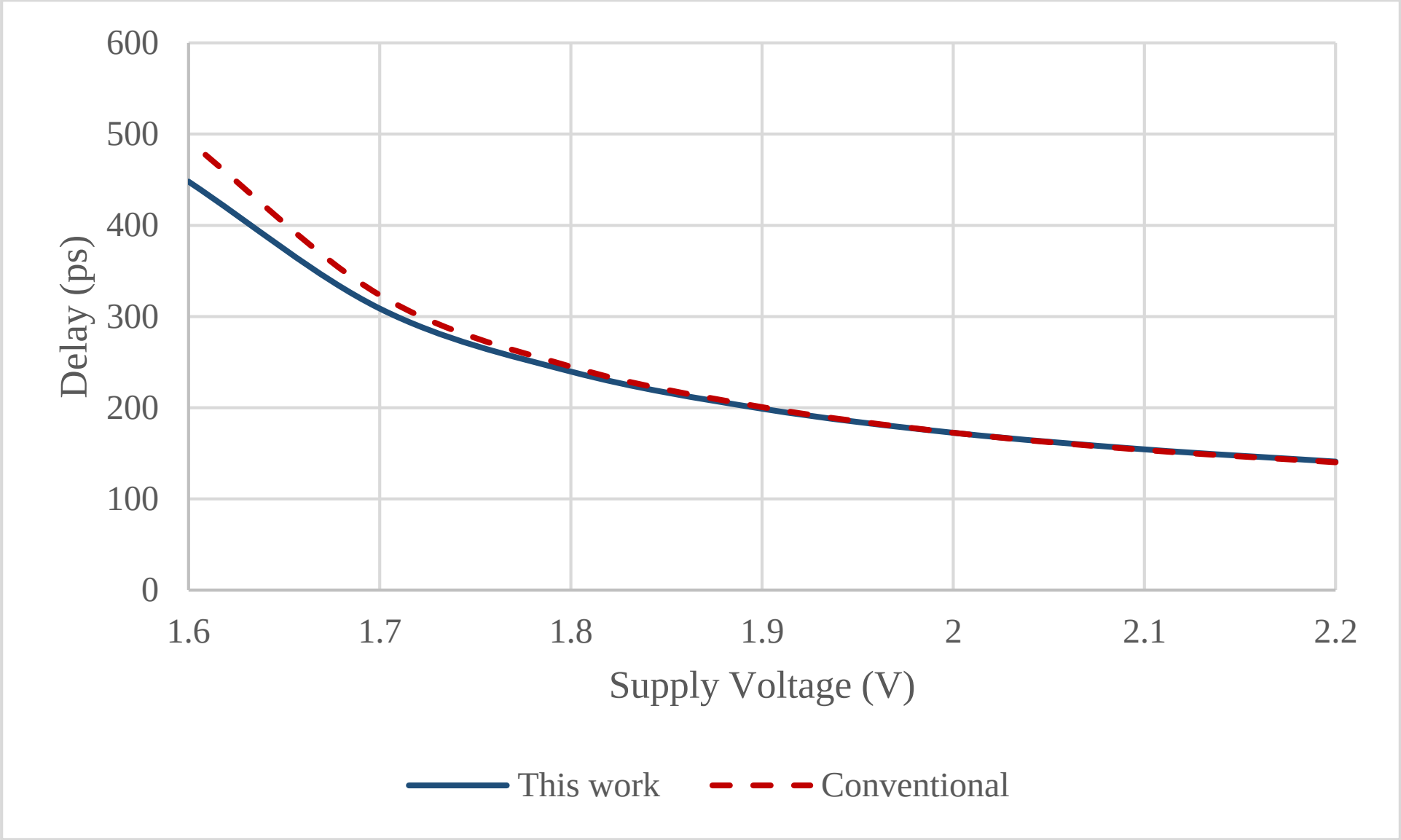

*Figure 14. Delay for different supply voltage of the proposed and conventional comparator.*

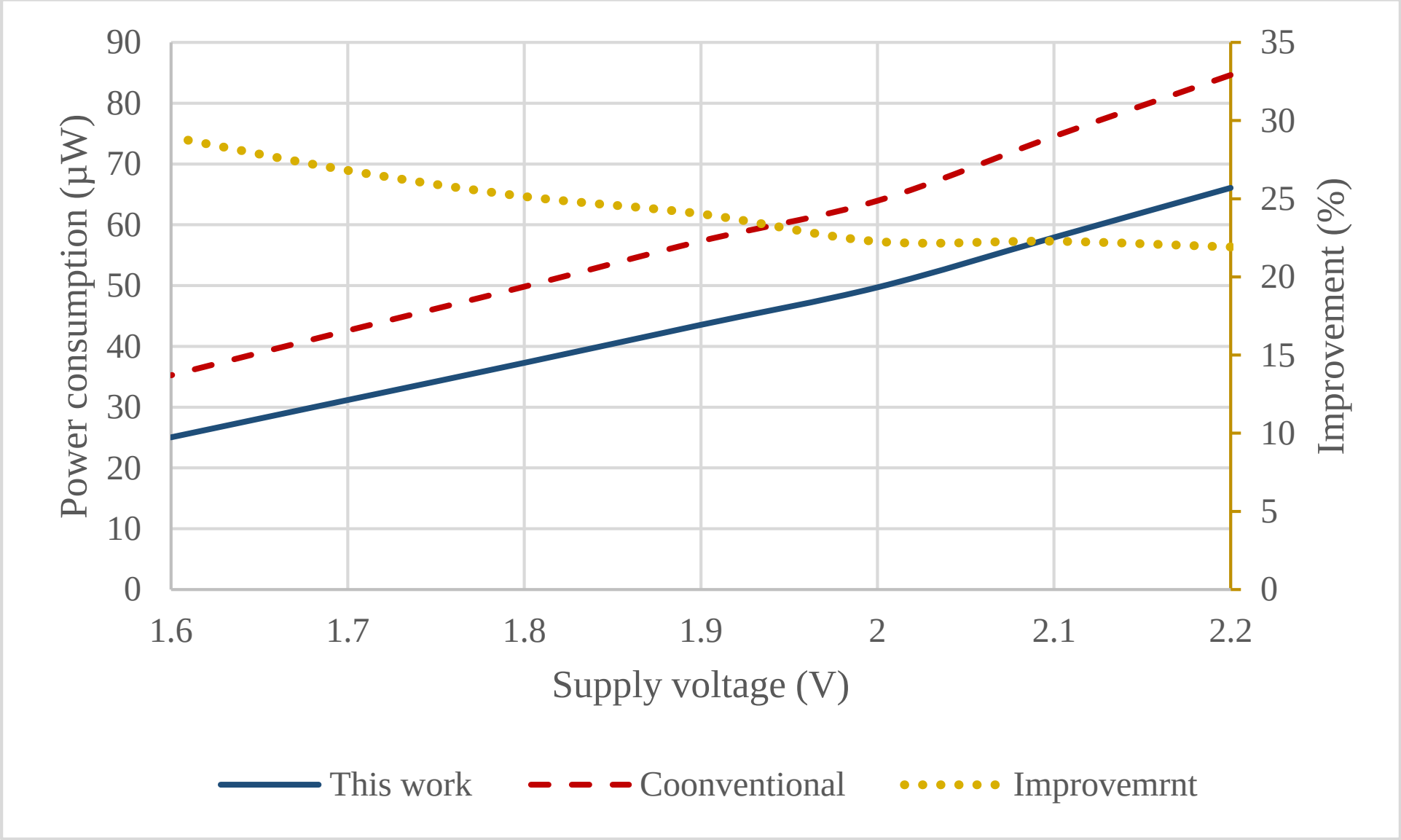

*Figure 15. Power consumption for different supply voltages of the proposed and conventional comparator.*

The delay and power consumption of the two schemes when the operation temperature varies from -20°C to 100°C are shown in Figure 16 and Figure 17 respectively. These figures show that in all the operation temperature range, the proposed structure is slightly faster and the power consumption is reduced in a rather flat manner.

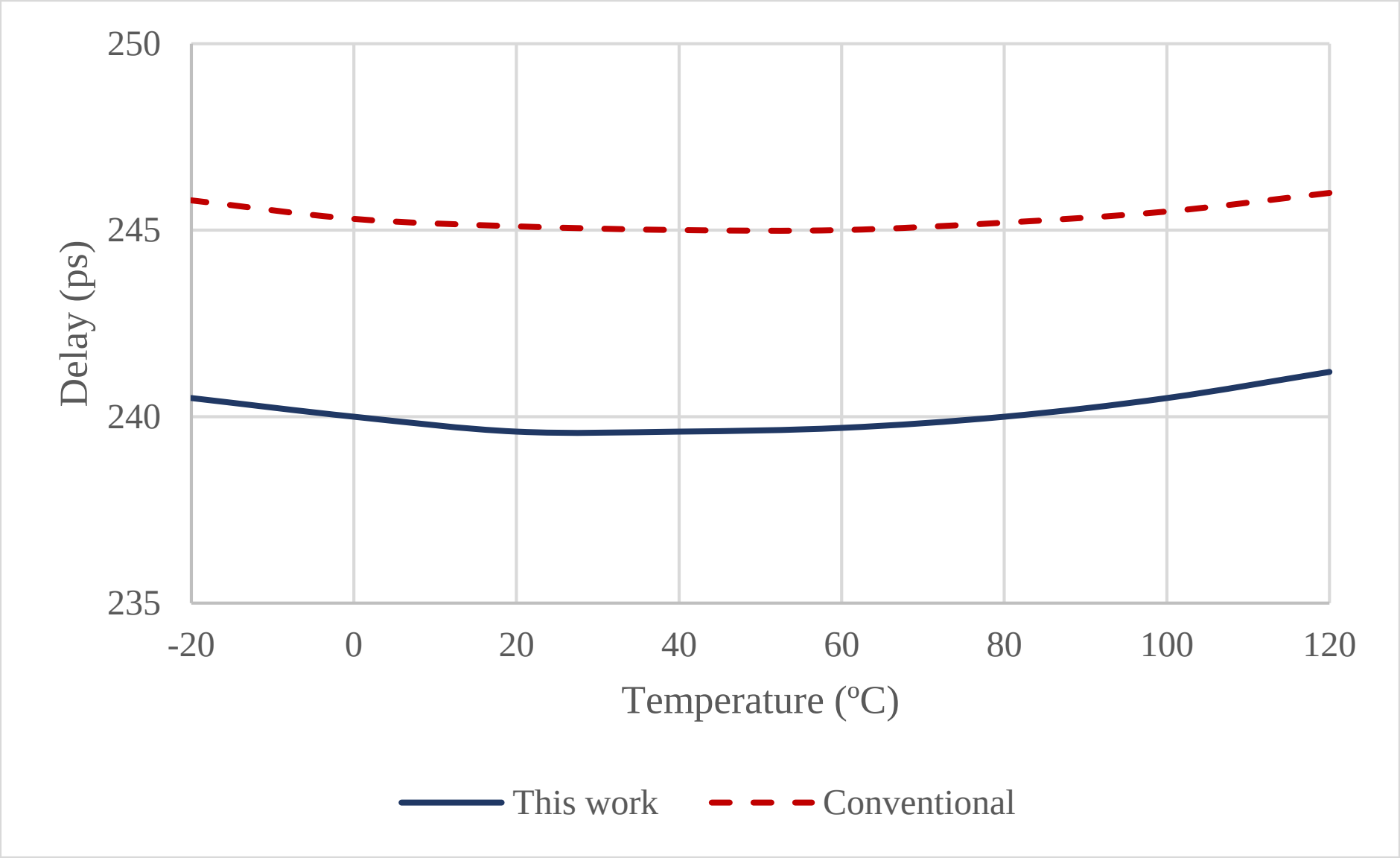

*Figure 16. Delay as a function of temperature of the proposed and conventional comparator.*

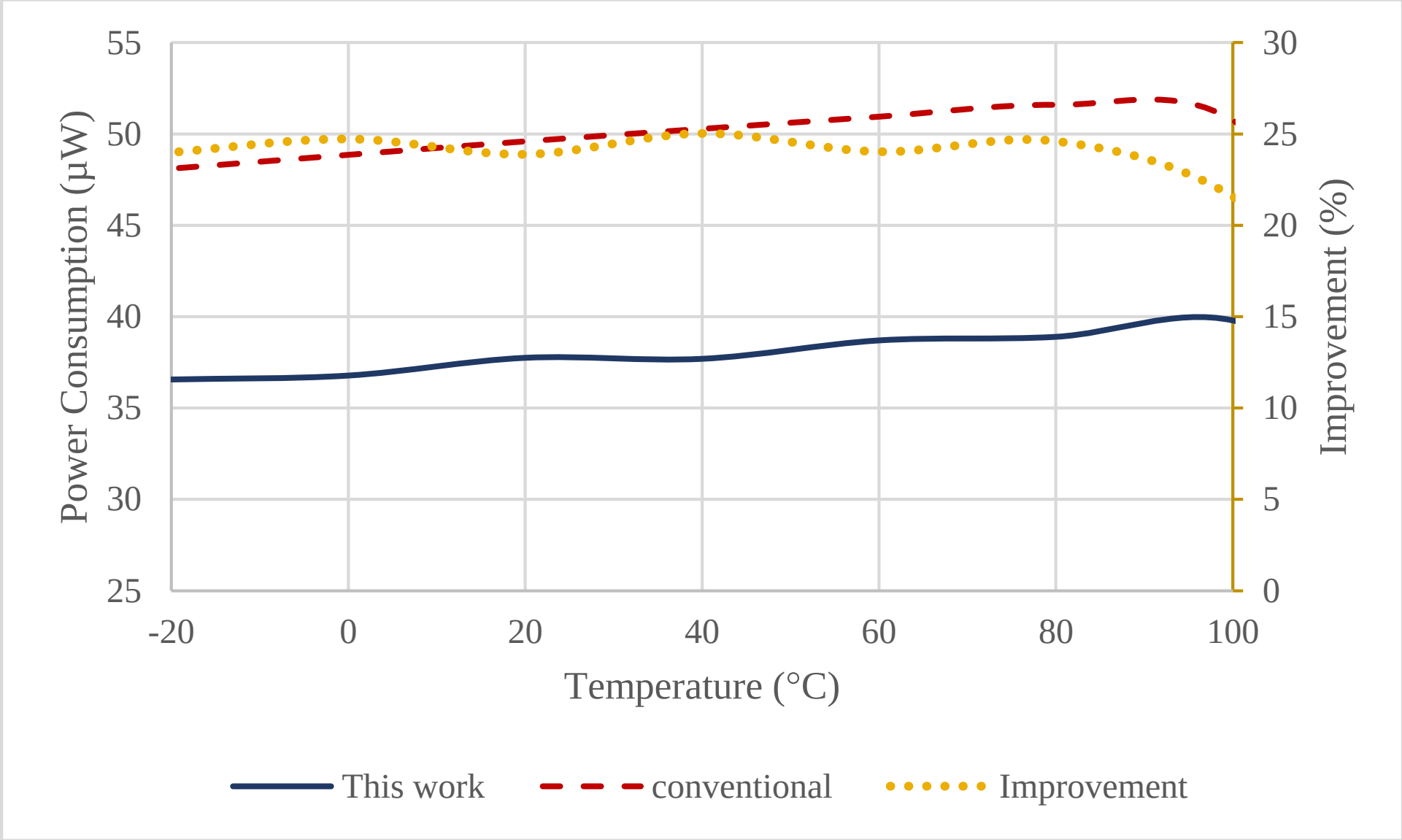

*Figure 17. Power consumption as a function of temperature of the proposed and conventional comparator.*

In order to verify the performance of the proposed offset cancelation design in the presence of device mismatch, Monte-Carlo analysis was conducted for 500 runs. The results of input referred offset for the proposed design before and after offset cancellation phase are shown in Figure 18 and Figure 19 respectively. The standard deviation and the average of the input referred offset before offset cancellation phase are 11.35mV and 3.48mV respectively. The standard deviation and the average of the input referred offset after offset cancellation phase become 0.620mV and 0.070mV respectively.

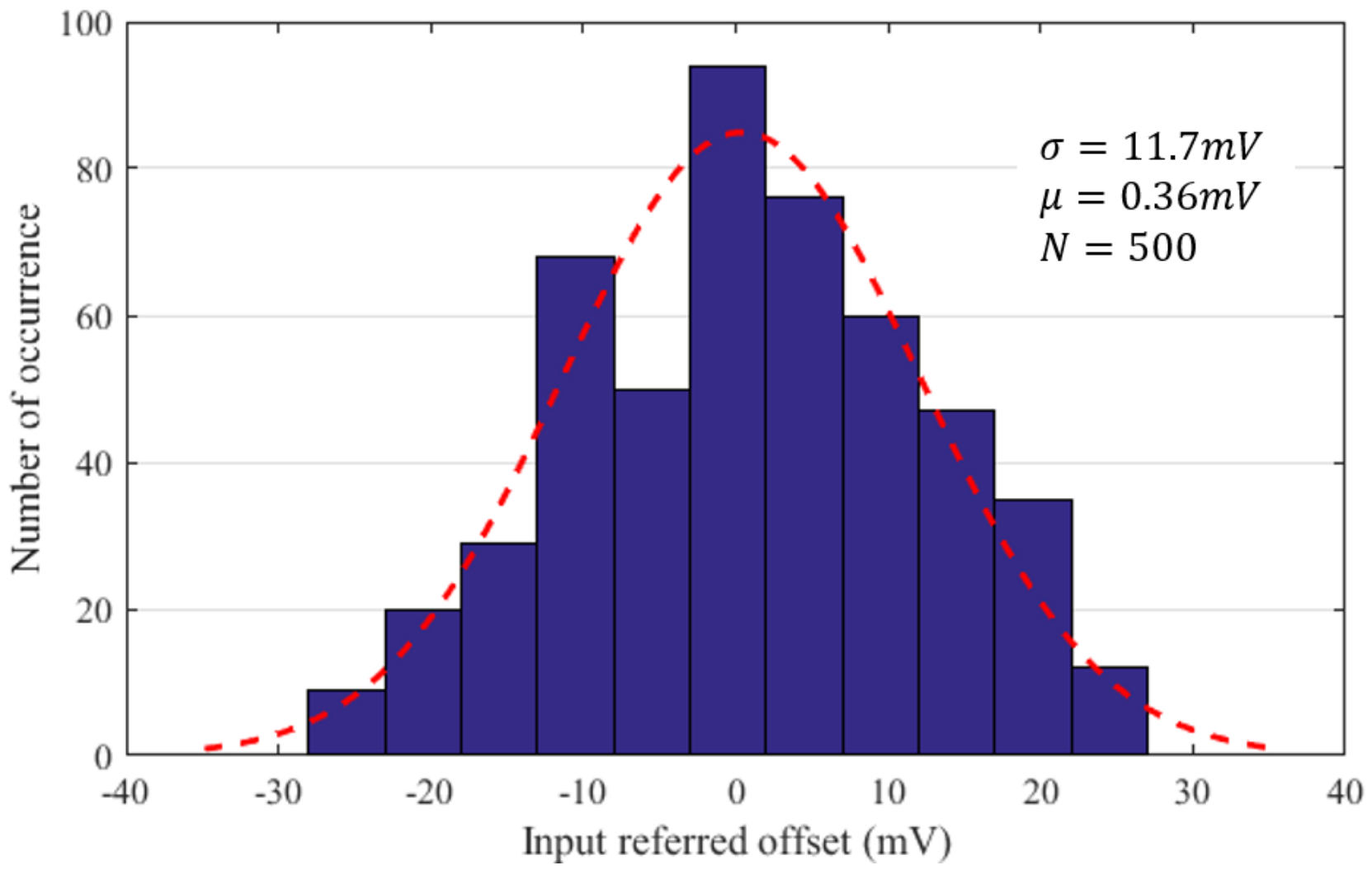

*Figure 18. Input referred offset, obtained with Monte-Carlo analysis for 500 iterations before offset cancellation.*

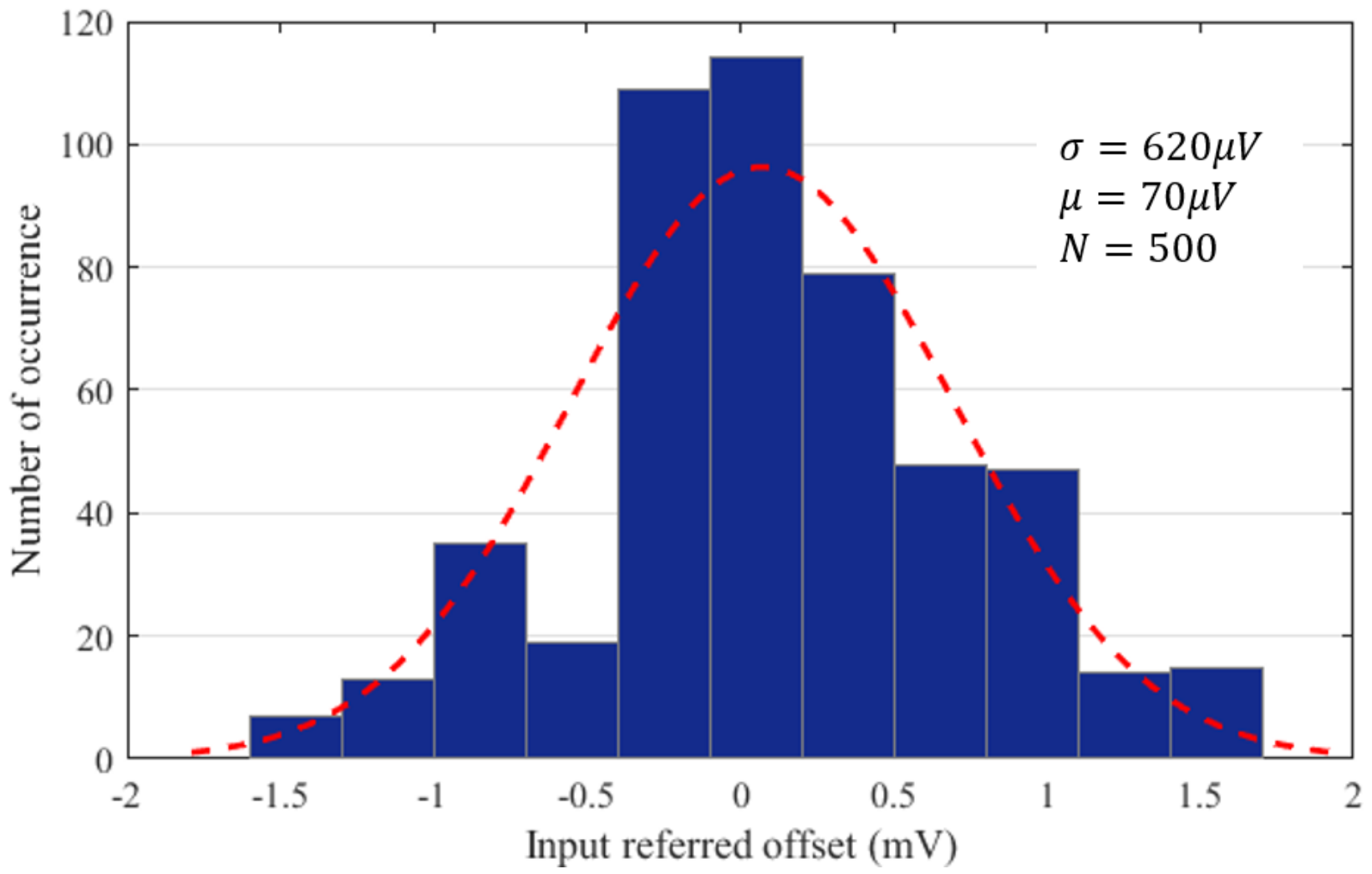

*Figure 19 Input referred offset, obtained with Monte-Carlo analysis for 500 iterations after offset cancellation.*

Figure 20, illustrates the layout of the introduced comparator. The introduced method contains analog and digital sub blocks for the offset cancellation phase. Because for the digital gates standard cells are used, the layout presented here consists of the analog parts.

Post layout simulation results for power consumption and delay of the proposed comparator against input common mode voltage and input differential voltages are shown in Figure 21 and Figure 22 respectively.

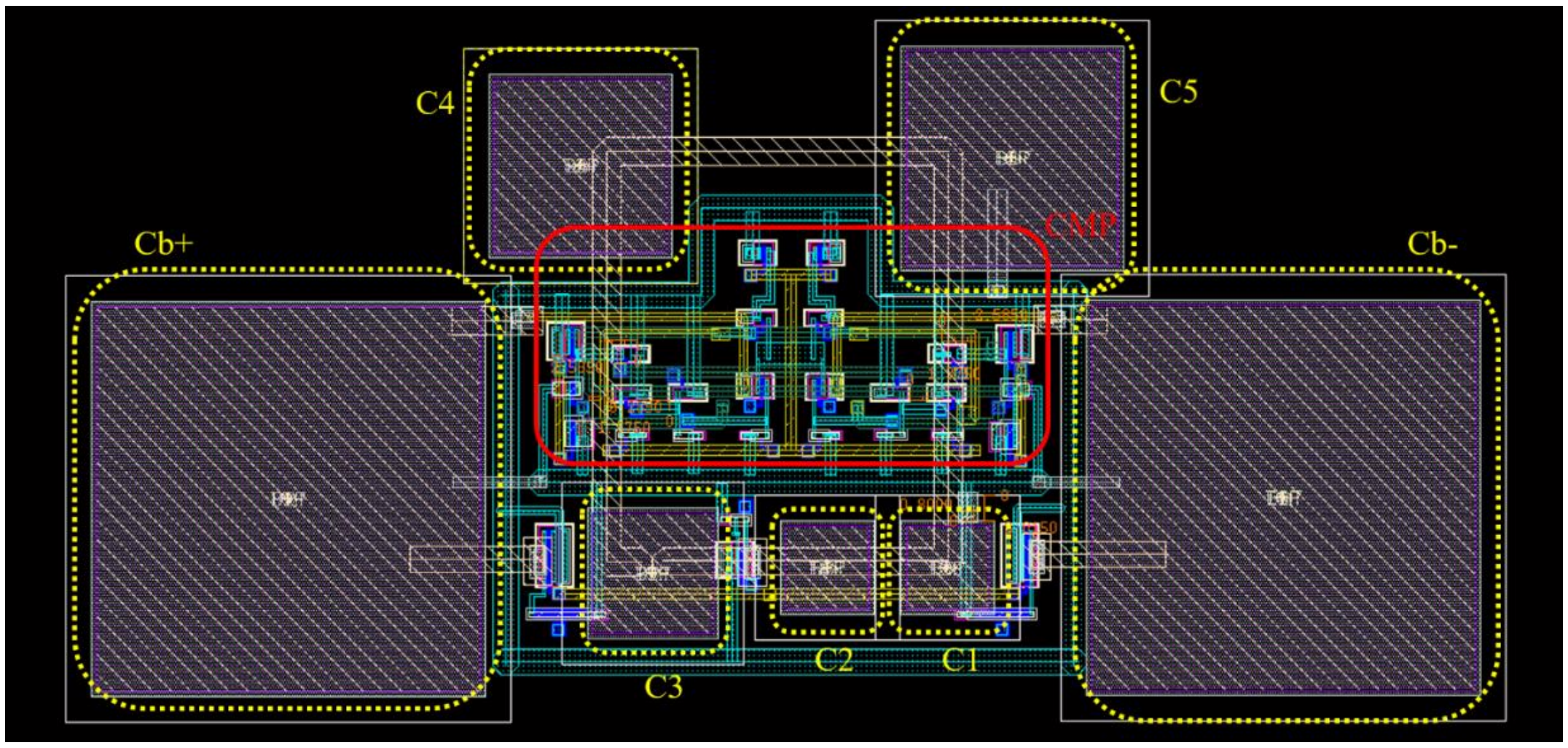

*Figure 20. Layout of the proposed comparator analog section.*

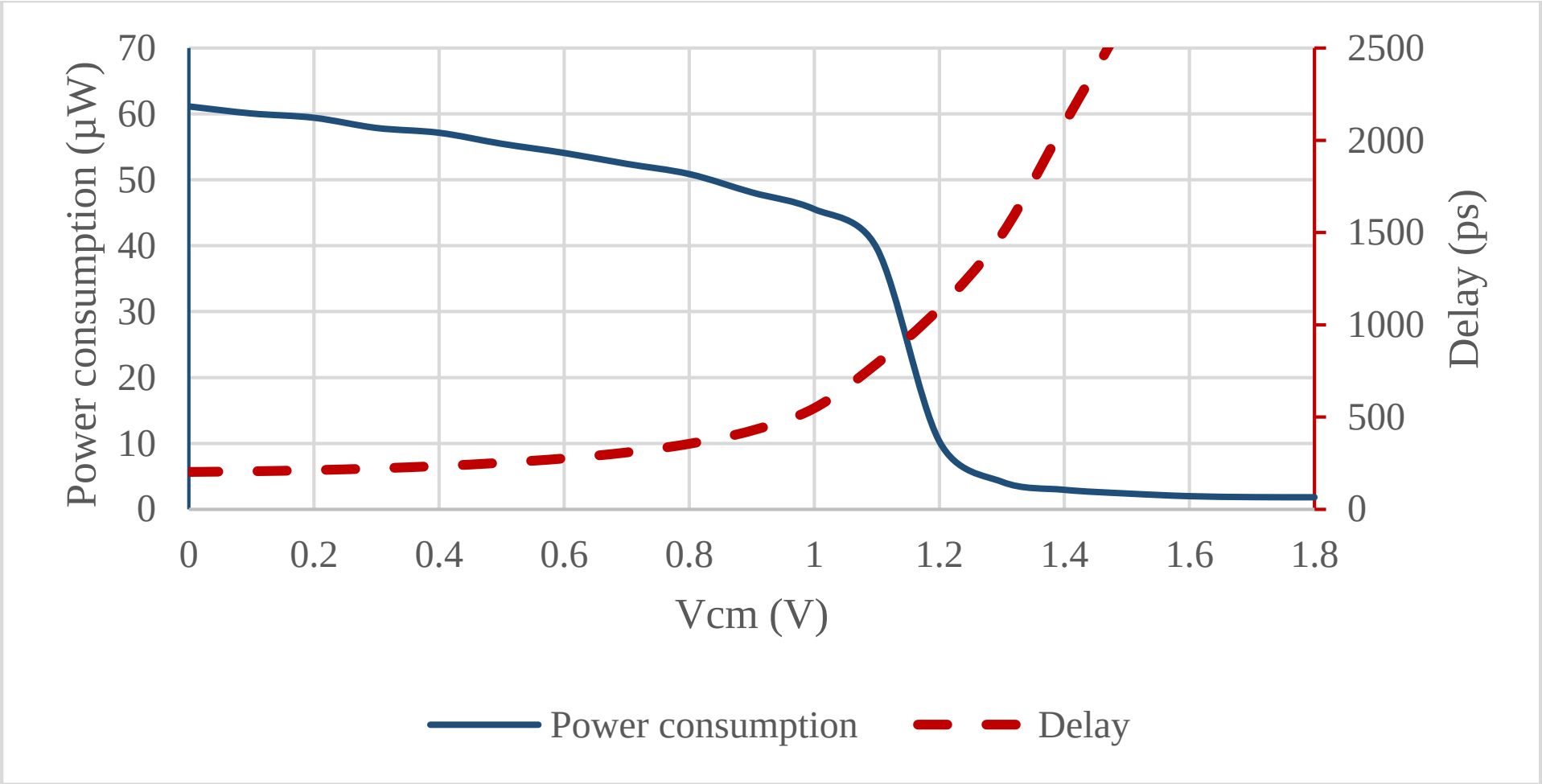

*Figure 21. Post layout simulation of power consumption and delay of the proposed comparator vs input common mode voltage.*

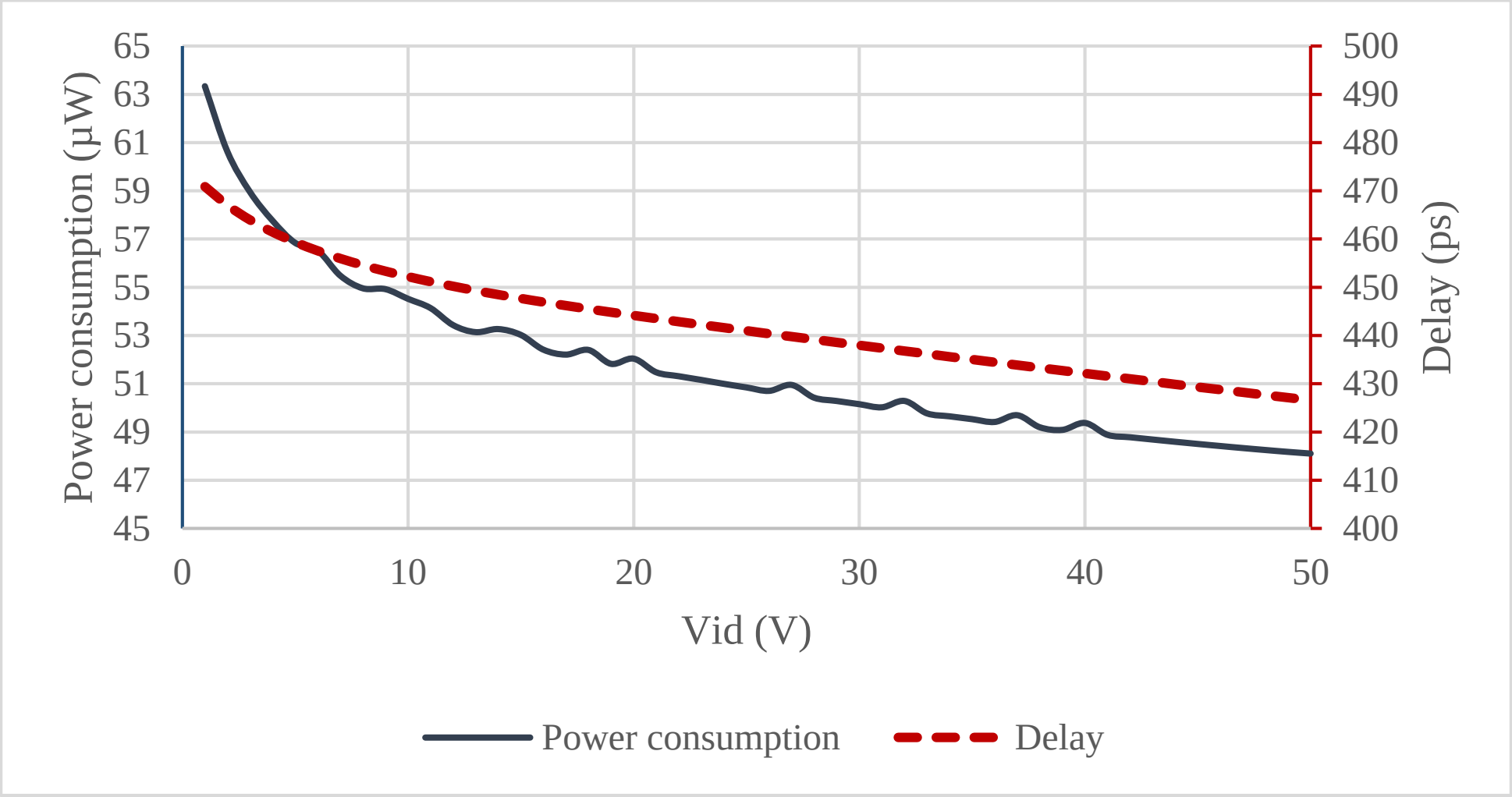

*Figure 22. Post layout simulation of power consumption and delay of the proposed comparator vs input differential voltage.*

The performance results of this study are compared with some relevant recent papers in Table 2. The results show that for comparable delay and the same simulation conditions, the lowest power consumption is obtained in the proposed solution. Besides, regarding the maximum clock frequency, the comparator presented in this study has the lowest output voltage delay time.

*Table 2. Performance comparison of this study with relevant recently published papers.*

| | **This work** | **Simulated results of [16]** | **[16]** | **[17]** | **[18]** | **[19]** | **[20]** |
|---|---|---|---|---|---|---|---|
| Technology | **180nm** | 180nm | 500nm | 180nm | 180nm | 180nm | 180nm |
| Supply voltage | **1.8V** | 1.8V | 5V | 1.8V | 1.8V | 1.8V | 1.8V |
| Power | **47 µW @Vid=1mV, F=500MHz** | 60 µW @ Vid=1mV, F=500MHz | 4.65 µW @ 200KHz | ≈200 µW @ Vid=1mV, F=500MHz | 420 µW @ Vid=1mV, F=500MHz | 347 µW @Vid=1mV, F=500MHz | 351 nW @ 200KHz |
| Delay | **256ps @ Vid=1mV** | 258ps @ Vid=1mV | NA | ≈450ps @ Vid=1mV | ≈350ps @ Vid=1mV | ≈400ps @ Vid=1mV | NA |
| Maximum clock frequency | **1GHz @Vid=1mV** | 1GHz @Vid=1mV | 33.3MHz | 500MHz | 4540MHz | 500MHz | 833MHz |
| Input referred offset ($\sigma$) | **628µV** | 9mV | 50.57µV (With OC) | 2.5mV | 2.5mV | 2.19mV | ≈1mV @Vic=.09V |
| **Area** | **5100($\mu m^2$)** | NA | 64K($\mu m^2$) | 315($\mu m^2$) | 453($\mu m^2$) | 361($\mu m^2$) | NA |

## 4. CONCLUSION

In ultra-low power SAR ADC designs, the comparator consumes a significant amount of power in comparison with other sub blocks [23]. In this paper, an effective power consumption reduction technique was proposed to reduce the overall power usage in dynamic comparators. The proposed technique offers lower power consumption with approximately same delay time. The power reduction technique was applied to a well-known low power comparator and was able to reduce its power consumption by 21.7% in the worst case, while leaving the delay time relatively intact. The overall power consumption of the proposed comparator is 47µW at 500MHz frequency. The comparator power and delay are simulated with different process corners, supply voltages and temperatures. A complete comparison was made with previous related studies to show the effectiveness of the proposed solution in the same operating conditions.

Since small dimensions of the proposed circuit imposed relatively high input referred offset to the main comparator, a time domain offset cancellation technique was also exploited. The offset cancellation technique uses the body voltages of the input devices to compensate the offset so that it does not affect the delay of the main comparator. Since the refresh rate of the offset cancellation technique is much lower than the operating frequency of the main comparator, its power consumption overhead is negligible.